\documentclass[aps,prb,groupedaddress,showpacs,twocolumn,floatfix]{revtex4-1}
\usepackage{mathbbol,amsthm,amsmath,amsfonts,amssymb,bm,xcolor,graphicx,psfrag,array,mathtools,lipsum,mathrsfs,comment,adjustbox,stmaryrd,braket}

\providecommand{\ignore}[1]{}
\providecommand{\aucmnt}[1]{#1}
\renewcommand{\aucmnt}[1]{}

\newcommand{\Comment}[1]{}

\begin{document}

 \title{A Theorem on Extensive Spectral Degeneracy \\ for Systems with Higher Symmetries  in General Dimensions}
 
\author{Zohar Nussinov}
\email{zohar@wustl.edu}
\affiliation{Rudolf Peierls Centre for Theoretical Physics, University of Oxford, Oxford OX1 3PU, United Kingdom}
\affiliation{LPTMC, CNRS-UMR 7600, Sorbonne Universite, 4 Place Jussieu, 75252 Paris cedex 05, France}
\affiliation{Department of Physics, Washington University, St.
Louis, MO 63160, USA}
\author{Gerardo Ortiz}
\affiliation{Department of Physics, Indiana University, Bloomington, IN 47405,USA}

\date{\today}

\begin{abstract}
We establish, in the spirit of the Lieb-Schultz-Mattis theorem, lower bounds on the spectral degeneracy of quantum systems with higher (Gauge Like) symmetries with rather generic physical 
boundary conditions in an arbitrary number of spatial dimensions. Contrary to applying twists or equivalent adiabatic operations, we exploit the effects of modified boundary conditions. 
When a general choice of boundary geometry is immaterial in approaching the 
thermodynamic limit, systems that exhibit non-commuting Gauge Like symmetries, such as the orbital compass model, must have an exponential (in the size of the boundary) 
degeneracy {\it{of each of their spectral levels}}. 
We briefly discuss why, in spite of the proven large degeneracy associated with infrared-ultraviolet mixing, some systems may still exhibit conventional physical behaviors, i.e., of those of systems with non-extensive degeneracies, due to entropic ``order by disorder'' type effects. 
\end{abstract}

\maketitle
 
\section{Introduction}

In the current work, we extend earlier results concerning the degeneracy of quantum systems (principally, those relying on the Lieb-Schultz-Mattis theorem \cite{LSM} and its higher dimensional generalizations \cite{Oshikawa,Matt,DMH}) to models with ``higher symmetries'' exhibiting an exponential (in the linear ($d=1$) system size, area ($d=2$), or higher ($d>2$) dimensional volume) spectral degeneracy \cite{benoit,DMH,NBCB,Ma,BCN,NF,DBM,BN,NOC1,NO,rev-compass}. Known arguments for establishing degeneracies (including, principally, the Lieb-Schultz-Mattis theorem) employ geometries with generalized Bloch type boundary conditions \cite{MPolarization}. While, as we will explain, the proof of exponential degeneracy for systems exhibiting higher symmetries is rather trivial in the classical limit, this is not the case for quantum systems.

In our efforts to establish the exponential spectral degeneracy that these quantum systems feature, we will follow an approach different from that of the known proofs of the Lieb-Schultz-Mattis theorem and its extensions. Our proof will, instead, rely on altering the boundary geometry of the system from that of a pristine hypercubic (or other) lattice with (Born-von Karman or Bloch type \cite{AM}) periodic boundary conditions to the very same lattice having an exterior boundary that is not cleaved so perfectly as to be symmetric along all Cartesian directions. For such general ``real life'' boundaries, we will rigorously establish a degeneracy that is exponential in the size of the boundary. As we will spell out in some detail, this modified boundary geometry may be 
changed by the inclusion of terms forming a ``zipper Hamiltonian''. When added to the existing interactions (i.e., the system with generic non-Bloch type boundary conditions for which we may rigorously prove exponential degeneracy in a finite-size system), the zipper Hamiltonian effectively restores conventional periodic boundary conditions. We will illustrate that, under rather mild assumptions, in the thermodynamic limit of various systems, the zipper Hamiltonian may not lift the exponential degeneracy. 

 Recently, there has been a renewed surge of interest in physical theories whose extensive degeneracy is intimately related to associated higher symmetries (also heavily studied in myriad contexts as ``$d$-dimensional Gauge Like Symmetries'' \cite{BN,NOC1,NO,Long-TQO,PNAS,PRLduality,AIP,rev-compass} ``unusual'' \cite{Harris}, ``stratified'' \cite{NBCB,BCN}, low-dimensional \cite{Ma,Arun}, ``sliding'' \cite{LF,NF}, ``infinite but nonextensive set of conservation laws'' \cite{XM}, ``generalized'', ``generalized global'' or ``higher form symmetries'' \cite{McGreevyreview,gaiotto,banks,son,QRH,Natis,johnm} (and their ``higher group'' mixtures \cite{Benini,Cordova}), and ``subdimensional'' or ``subsystem symmetries'' \cite{slagle-field-theory,You,Stephen,Shirley3,tri2020,hosho,cao}). Numerous finer classifications of these symmetries and their extensions exist (e.g., whether they are invertible or not) with non-trivial consequences, e.g., \cite{invert1,invert2,invert3,invert4}. References [\onlinecite{Long-TQO}] and [\onlinecite{PNAS}] initiated a study explaining how the gauge like character of these symmetries mandates, in various circumstances, topological order. Subsequent, very penetrating, results linking these symmetries to topological phases appeared in Ref. [\onlinecite{gaiotto}] and other illuminating works. As befits their name(s), these higher or $d$-dimensional Gauge Like Symmetries act nontrivially on a $d$ (or $d+1)$-dimensional spatial (or space-time) subvolume of a theory  defined in $D$ (or $D+1$) spatial (space-time) dimensions. The $d=0$ and $d=D$ (or $d=D+1$) dimensional spatial (or space-time) limits correspond to the standard local (gauge) and global symmetries. 

The more non-trivial $d$-dimensional Gauge Like Symmetries lie between the diametrically opposite limits of local and global symmetries ($0<d<D$) (and thus act on a lower dimensional subvolume of the physical system). A general study of the consequences of these symmetries and, in particular, of the dimensional reductions that they imply first appeared in \cite{BN,NOC1,NO,Long-TQO,PNAS,PRLduality,AIP}. The upshot is that, as a rule, when the dimension of these symmetries is smaller than that of the system ($d <D$), a generalization \cite{BN,NOC1} of Elitzur's theorem \cite{Elitzur} (physically capturing the existence of $d$-dimensional topological defects in $D$ spatial dimensions) leads to an effective dimensional reduction from $D \to d$ dimensions for various observables and general spatio-temporal correlation functions \cite{BN,NOC1,NO,Long-TQO,PNAS,PRLduality,AIP,rev-compass}. For low $d$, the  proliferation of $d$-dimensional topological defects at finite temperatures may, unfortunately, lead to a loss of memory in topological quantum memory schemes \cite{NO}. This particular corollary may be proven by application of the generalized dimensional reduction inequalities \cite{NOC1,NO,Long-TQO,PNAS,PRLduality,AIP} to autocorrelation functions. For the special cases of stabilizer models (including the celebrated  Kitaev Toric Code \cite{TORIC}, the Chamon model \cite{chamon}, the Haah Code \cite{Haah},  the X-cube \cite{PhysRevB.94.235157}, and other models, e.g.,  \cite{NO,zack,zack1}) the lower dimensional symmetry inequalities are further augmented by exact dualities  \cite{NOC1,NO,Long-TQO,PNAS,PRLduality,AIP,zack,zack1} that map the partition functions and the equations of motion of these higher 
($D>1$) dimensional systems to those of dual one-dimensional systems. 
A trivial consequence of those dimension reducing dualities is that the equations of motion for general observables will exactly map to those of the lower dimensional dual systems \cite{NOC1,PRLduality,AIP,lowdeqsmotion}. This implies that their autocorrelations are identical to those of lower dimensional systems, suggesting that some of these systems might not be immune to thermal fluctuations \cite{lowdt2}.

Intrinsically, the action of $d$-dimensional symmetries on a spatially smooth low-energy configuration (i.e., one with infrared (IR) Fourier components) may generate other degenerate low-energy states that are not, at all, necessarily slowly varying in space, and have significant weight associated with their ultraviolet (UV) components. This seemingly rather odd facet has reignited various questions and investigations as to how continuum field theories may describe such highly degenerate systems with unconventional IR-UV mixing \cite{slagle-field-theory,seiberg2021,Natis,Pra}. Of particular current note are studies of ``fracton'' theories \cite{Pra,chamon,QRH,Bravyi,Haah,PhysRevB.95.155133,PhysRevB.94.235157,Natis,Nan1,Halasz,Shirley1,Shirley2,Shirley3,Shirley4,Shirley5,Slagle1,elastic-fracton,devakul,You,son,Sid2,Slagle17,zack,zack1} and their hybrids \cite{hybrid1,hybrid2} (including, using the above noted dualities, the exact finite temperature solutions \cite{NOC1,zack,zack1} of all of the first fracton models \cite{chamon,Bravyi,Haah,PhysRevB.95.155133}). Some qualitatively similar behaviors also appear in  theories exhibiting a ``fragmentation'' of their Hilbert space into individual ergodic subspaces that do not readily enable transitions from one subspace to another \cite{sala}. Predecessors of current fracton-like models featuring an extensive number of $d$-dimensional Gauge Like symmetries have been the ``compass models'' that we will study in Section \ref{sec:POC} \cite{NBCB,BCN,BN,Ma,NF,DBM,NO,NOC1,rev-compass,Long-TQO,PNAS,PRLduality,AIP,KK,jan1,120compass,simon}, which include celebrated Kugel-Khomskii type models \cite{rev-compass,KK,Harris}. These decades-old models \cite{KK} and their extensions describe orbital (and spin) degrees of freedom in transition metal compounds \cite{KK,jan1,120compass}. Compass models are some of the simplest models capturing the quintessential physics associated with the physical connections between degeneracies and $d$-dimensional Gauge Like Symmetries that we wish to explore. 

 Before proceeding further, we briefly comment on the relation between symmetries and degeneracies in the classical limit with a particular focus on theories exhibiting higher symmetries such as the classical variant of the 90$^{\circ}$ square lattice compass model \cite{rev-compass} that we will specifically elaborate on later. In classical theories, no entanglement exists between the disjoint spatial regions where the different higher symmetries may operate. Thus, given a classical (spin, field, or other) configuration ${\cal{C}}$, one may turn, in a binary fashion, ``on'' or ``off'' the ${\cal{M}}_{\sf cl}$ independent classical higher symmetries that operate on different subregions of ${\cal{C}}$. The $2^{{\cal{M}}_{\sf cl}}$ different binary strings defined by the choice of applying/not applying these ${\cal{M}}_{\sf cl}$ different symmetries may, generally, be associated with $2^{{\cal{M}}_{\sf cl}}$ states that are degenerate with ${\cal{C}}$.  Symmetries that have more than two group elements, allowing for more than just ``on'' or ``off'' applications, may yield a degeneracy that is larger yet.
 If the number of independent symmetries ${\cal{M}}_{\sf cl}$ scales as a lower dimensional subvolume of the system then the above implies sub-extensive exponential degeneracies. Apart from compass type systems, different subvolume exponential degeneracies  were also found to appear in classical field theories with non-Abelian backgrounds \cite{NAG}, originally introduced as models of glasses \cite{SN}, as well as a host of classical spin models with spiral ground states \cite{spiral1,spirals}. In both classical and quantum continuum gauge theory formulations that treat elasticity as that of a strongly anisotropic medium in space-time, defects are constrained to lower dimensional regions \cite{elastic1,elastic2} matching more recent fracton inspired gauge theory formulations \cite{elastic-fracton}. 

As is well known, in quantum theories the relation between symmetries and degeneracies is more intricate. Quantum fluctuations may lift classical degeneracies.
These fluctuations can select an ordered state out
of a plethora of many classical degenerate states -- a mechanism often colloquially called ``quantum order by disorder'' \cite{rev-compass,QOBD,QOBD0,QOBD1,QOBD2,QOBD3,QOBD4}.
This phenomenon is superficially reminiscent of anomalies in quantum field theories in the following sense. Although classical $d$-dimensional symmetries may remain symmetries of the quantum theory, their degeneracies  need not carry over to their quantum counterparts. Indeed, in the quantum arena, symmetries {\it do not} straightforwardly imply the existence of degeneracies (e.g.,  eigenstates of the Hamiltonian may transform as singlets under the various symmetry operations). One of the simplest textbook examples illustrating this dictum is afforded by the absence of non-trivial transformations of the (even and odd) symmetric ($D=1$) double well eigenstates under the parity symmetry operator. By sharp contrast, degeneracies always necessarily mandate the existence of symmetries (within any linear subspace spanned by $n \ge 2$ orthogonal degenerate eigenstates, there is an internal $SU(n)$ ``rotation'' symmetry associated with general superpositions of these degenerate states) that, rather trivially, does not change the energy (see, e.g., Ref. [\onlinecite{Long-TQO}])). Determining the spectral degeneracies is typically done on a case by case basis for numerous problems across diverse fields, both fundamental and applied \cite{Eckart,Wigner1959,Weyl,Morton,Tinkham}. Degeneracies that do not exist in finite size systems can, in some instances, rear their head so as to only emerge asymptotically in the thermodynamic limit. The latter situation arises in diverse situations including, e.g., those cases in which tunneling between edge states  is suppressed in the thermodynamic limit \cite{sagi}. 
In the context of the compass model that we will use as an example in this work, an exponential degeneracy was suggested to appear in the thermodynamic limit in Refs. \cite{benoit,DBM}.

Our work aims to show how, in specific circumstances, higher symmetries may rigorously lead to degeneracies in quantum systems. 
Furthermore, if the choice of boundary conditions is immaterial spectral degeneracies may still emerge in the thermodynamic limit. 

We will discuss how {\it localized} changes (zipper interaction terms) in the system Hamiltonian may connect it to a system with exactly provable degenerate states.

Our principal findings can be succinctly summarized by two basic inter-related maxims for a system of linear size $L$ featuring $d$-dimensional Gauge Like Symmetries:

{\bf (A)} Symmetries and, in particular, $d$-dimensional Gauge Like Symmetries mandate exponential degeneracies when the choice of boundary conditions is immaterial in the thermodynamic limit. That is, the logarithm of the 
degeneracy of {\it each level} \cite{explain-A} of energy $E$, $g(E)$, scales asymptotically at least as fast as $L^{d'}$, 
\begin{eqnarray}
\lim_{L \to \infty} L^{-d'} \ln g(E) >0 ,
\end{eqnarray}
with 
\begin{eqnarray}
\label{codim}
d'=D-d 
\end{eqnarray}
denoting the co-dimension of the spatial dimension ($d$) of the higher (Gauge Like) symmetries. 
 
 {\bf (B)} Given boundary conditions for which exponential degeneracy may be rigorously proven for finite system sizes, the same exponential degeneracy may persist in the thermodynamic limit in the presence of other boundary conditions or interactions with different environments. Unlike {\bf (A)}, some arguments underlying this maxim are not as strong.
 
\section{On Exponential Degeneracy and Boundary Conditions}
\label{sec:exp}

\subsection{A Fundamental Theorem}

In order to establish maxim {\bf{(A)}}, we start with a simple Lemma.

{\bf{Lemma 1.}}
Consider a system governed by a Hamiltonian $H_{\sf open}$ on a $D$-dimensional spatial volume $\Omega_{D}$, and free (open) boundaries, for which there are two ``dual'' sets of independent symmetries $\{U_{a}\}$ and $\{V_{a}\}$ (with $a= 1,2, \ldots  {\cal{M}}$) satisfying the following conditions:

{\bf{(1)}} All operators in one of these two sets mutually commute with one another,
\begin{eqnarray}
\label{trivcomm}
[U_{a}, U_{a'}] =0.
\end{eqnarray}

{\bf{(2)}} For any symmetry $U_{a}$ in the above set there is only a single dual operator $V_{a}$ that does not commute with it. Specifically, 
\begin{eqnarray}
\label{aa}
[U_{a}, V_{a}]  \equiv W_{a} \neq 0  \mbox{ and } [U_{a}, V_{a'}] =0 ~ \mbox{ for } a \neq a',
\end{eqnarray}
where the operator $W_{a}$ does not have a null space. 

When the above conditions are met, each eigenstate of $H_{\sf open}$ is, at least, $2^{{\cal{M}}}$ fold degenerate. 

{\it Proof.}
Given condition {\bf{(1)}}, we may label all of the eigenstates of $H_{\sf open}$ by the eigenvalues $\{\lambda_{a}\}$ of the $ {\cal{M}}$ independent symmetries $\{U_{a}\}$ (along with any additional quantum numbers if additional degeneracies appear in a sector of fixed eigenvalues of all $\{U_{a}\}$ operators). That is, all of the energy eigenstates are of the form 
\begin{eqnarray}
\label{el-state}
| \psi \rangle = |\lambda_{1} \ldots \lambda_{a} \ldots \lambda_{{\cal{M}}}, \{\nu\} \rangle,
\end{eqnarray}
with $\{ \nu \}$ additional labels (comprised of the energy eigenvalue and other quantum numbers if any) for all states within a given sector of $\{\lambda_{a}\}_{a=1}^{\cal{M}}$. 
To illustrate that for any $a$, there are, at least, two states with different eigenvalues of $U_{a}$ having the same energy, we may next invoke condition {\bf{(2)}}. As the commutator $W_{a}$ does not have a null space ($W_{a} | \psi \rangle \neq 0$), it follows that $| \psi \rangle$ is not an eigenstate of $V_{a}$. Thus, the state $V_{a} | \psi \rangle$ is linearly independent of $| \psi \rangle$. Since $V_{a}$ is a symmetry of the Hamiltonian, it further follows that the two states $| \psi \rangle$ and $V_{a} | \psi \rangle$ are linearly independent eigenstates of $H_{\sf open}$ sharing the same eigenvalue (energy). Lastly, the commutativity $ [U_{a}, V_{a'}] =0 ~ \mbox{ for } a \neq a'$ implies that $V_{a} | \psi \rangle$ is still an eigenstate of all other operators $U_{a'}$ with $a' \neq a$. Repeating the latter sequence of steps when the two states $ | \psi \rangle$ and $V_{a} | \psi \rangle$ are acted by other symmetries $V_{a'}$ with $a' \neq a$, it follows that the states
$V_{a}^{n_{a}} V_{a'}^{n_{a'}} | \psi \rangle$ (where $n_{a}=0,1$
and $n_{a'}=0,1$) constitute four linearly independent eigenstates of the Hamiltonian. 
Recursively iterating the above procedure for other symmetries of the $V$ type, one sees that the $2^{\cal{M}}$ states $V_{1}^{n_{1}} V_{2}^{n_2} \ldots V_{\cal{M}}^{n_{{\cal{M}}}} | \psi \rangle$ (where for each $1 \le a \le {\cal{M}}$, we may set $n_{a}$ to be either $0$ or $1$) are linearly independent degenerate eigenstates of the Hamiltonian. The existence of (at least) the above $2^{\cal{M}}$ independent degenerate eigenstates establishes the Lemma.  ~$\blacksquare$

This leads us to our central theorem, 
\bigskip

{\bf{Theorem 2. }}
General Hamiltonian systems (e.g., lattice theories on general graphs) that exhibit, at least,  
\begin{eqnarray}
\label{ML}
{\cal{M}} \ge c L^{d'}, \ c,d'>0, 
\end{eqnarray} 
independent symmetries that satisfy the conditions of Lemma 1, have for all energy eigenvalues $E$ of the Hamiltonian a degeneracy
\begin{eqnarray}
g(E) \ge 2^{c L^{d'}}. 
\end{eqnarray}
\bigskip

{\it Proof.} This follows from an immediate application of Lemma 1 given Eq. (\ref{ML}).
~ $\blacksquare$
\bigskip

Moreover, if the symmetries in Lemma 1 correspond to higher (or Gauge Like) symmetries of dimension $d$, then $d'=D-d$ as explained below.  
We note that since there may be additional multiplicities of $\{\nu\}$ associated with any set of the eigenvalues of the symmetry operators, the degeneracy of each of the eigenvalues $E$ of $H_{\sf open}$ will be an integer multiple of $2^{\cal{M}}$. 
We stress that this exact exponential degeneracy applies to all eigenstates and is not limited to the ground state sector of the system. Indeed, later on we will discuss finite temperature theories.

The above considerations may be sharpened as follows

{\bf Corollary 3.} 
Consider the situation in which (i) the operators $\{U_{a} \}$ and $\{V_{a} \}$ are {\it not symmetries} of the Hamiltonian and commute with $H_{\sf open}$ only in a given fixed energy 
$E$ and (ii) the operators $\{U_{a} \}$ and $\{V_{a}\}$ satisfy the conditions of Lemma 1. In this case, the proofs of Lemma 1 and Theorem 2 may be repeated for the projected Hamiltonian $P_{E} H_{\sf open} P_{E}$  with $P_{E}$ being the projection operator to the space of energy $E$ to establish exponential degeneracy. Thus, even if  $\{U_{a} \}$ and $\{V_{a} \}$ are not exact symmetries of the full Hamiltonian $H_{\sf open}$ and only commute with it within a given energy subspace (i.e., these operators are {\it emergent symmetries} \cite{BatistaOrtiz-2004}) then in that energy subspace the exponential degeneracy of Theorem 2 is ensured. Similar results apply if $\{U_{a} \}$ and $\{V_{a} \}$ satisfy the conditions of Lemma 1 and only become symmetries in those projected sectors of fixed quantum numbers other than those of constant energy.  
\bigskip

In the systems that we will investigate in more detail, the many-body Hamiltonian $H_{\sf open}$ is a sum of few body interactions or ``bonds'' $b_{\gamma}$,
\begin{eqnarray}
\label{eq:bond:def}
H_{\sf open} = \sum_{\gamma} b_{\gamma}.
\end{eqnarray}

We next outline the reason why, for {\it typical spin and bosonic systems}, boundary conditions can be invoked such that the conditions of Lemma 1 apply for ${\cal{M}} \ge c L^{d'}$ independent symmetries with the co-dimension $d'$ of Eq. (\ref{codim}). In fact, {\it common open boundary realizations of real physical systems satisfy Eq. (\ref{ML})}. In what follows, we will denote the generic local spin or Bose operators by $\{\phi^{\mu}_r\}$.
Here, $\mu$ is an internal label and $r$ the appropriate external position (site) index. These operators will explicitly become the Pauli operators $\{\sigma^\mu_r\}$ (with $\mu=x,y,z$) in spin $S=1/2$ models. For generic finite size lattices (or graphs) different from those with periodic boundary conditions, such an extensive number of independent symmetries trivially appear as a result of geometric considerations. In Bose and spin systems, whenever the symmetry operators $\{U_{a} \}$ have their support on $d$-dimensional regions $\{{\cal{R}}_{a}\}$ that are spatially disjoint (i.e., when these regions share no common sites), then they will commute since Bose and spin operators on different spatial sites trivially commute and Eq. (\ref{trivcomm}) is satisfied. This is generally not true when they do overlap and non-trivial commutators will appear. Indeed, if the regions ${\cal{R}}_{a}$ and $\tilde{{\cal{R}}}_{a}$ (respectively, the spatial supports of the symmetries $U_{a}$ and $V_{a}$) share common sites while  ${\cal{R}}_{a}$ and $\tilde{{\cal{R}}}_{b}$ (respectively, the spatial supports of the symmetries $U_{a}$ and $V_{b}$ with $a \neq b$) share no common sites then, generally, $U_{a}$ and $V_{a}$ need not commute with one another. Lastly, $U_{a}$ and $V_{b}$ with $a \neq b$ do not have overlapping spatial support and may trivially commute (giving rise to Eq. (\ref{aa})). In mathematical terms, 
\begin{eqnarray}
&& {\cal{R}}_{a} \cap {\cal{R}}_{b} = \emptyset , \mbox{for $a \neq b$}, \nonumber
\\ && {\cal{R}}_{a} \cap {\tilde{{\cal{R}}}}_{b} = \emptyset , ~ \mbox{for $a \neq b$}, \nonumber
\\ &&  {\cal{R}}_{a} \cap {\tilde{{\cal{R}}}}_{a} \neq \emptyset .
\end{eqnarray}
In many systems, geometry alone (whether the spatial (or spatio-temporal) support of the symmetry operators overlaps or does not) mandates Eqs. (\ref{trivcomm}) and (\ref{aa}). That is, geometry determines  whether the symmetry operators commute. By definition, for a system in $D$ spatial dimensions exhibiting $d$-dimensional Gauge Like Symmetries, the symmetries $U_{a}$ or $V_{a}$ have their support on a $d$-dimensional spatial region ${\cal{R}}_{a}$ and $\tilde{{\cal{R}}}_{a}$.
For a generic open volume $\Omega_{D}$ in $D$ dimensions (i.e., not one with square or cubic boundaries), there are  foliations of $\Omega_{D}$ into ${\cal{O}}(L^{d'})$ regions $\{{\cal{R}}_{a}\}$ and into ${\cal{O}}(L^{d'})$ regions $\{\tilde{{\cal{R}}_{b}}\}$ on which the symmetries $\{U_{a}\}$ and $\{V_{b}\}$ have their support. That is, for random open boundaries of $\Omega_{D}$, we can partially slice it into ${\cal{O}}(L^{d'})$ non-overlapping $d$-dimensional hyperplanes satisfying the conditions of Lemma 1. For fermionic (or non-commuting elementary degrees of freedom) systems, if the symmetries involve an even number of operators per site (and/or regions $\{{\cal{R}}_{a}\}$ and  $\{\tilde{{\cal{R}}_{b}}\}$ having an even number of sites) then the mutually commuting nature of the symmetries on different disjoint regions is, once again, ensured. 

The above discussion might seem a bit abstract. In order to make it more concrete, we will shortly turn (Section \ref{sec:POC}) to simple examples. An important ingredient will be the use of general boundary conditions that are different from those of the conventional Born-von Karman form. These general boundary conditions are not merely academic. Indeed, real materials are not wrapped around tori that endow them with periodic boundary conditions nor are they boxed to textbook type square (or hypercubic) boundaries. The boundaries $\partial \Omega_{D}$ of real physical systems $\Omega_{D}$ are typically very different from these conventions and may generally span many possible geometries. The use of periodic boundaries has indeed largely been a matter of convenience (i.e., since these allow for a Fourier space analysis and have no identifiable boundary sites so as to emulate bulk macroscopic systems in which nearly all real space sites lie far from the boundaries). The use of periodic boundary conditions has, for these and related reasons, become very common by now. In Section \ref{sec:cyl} below, we will examine boundaries different from the periodic ones to illustrate how our central Theorem 2 mandates an extensive degeneracy of this system. Nearly identical constructs may be introduced for many other theories and disparate boundary conditions different from those that are customarily employed. Before proceeding further however, we must note that our considerations do not apply 
for many other systems, such as Kitaev's Toric Code model \cite{TORIC}, where local operators are defined on bonds and not sites
(vertices). In particular, 
the ($D=2$ dimensional) Kitaev Toric Code model \cite{TORIC} supports its well-known $d=1$ symmetries \cite{Long-TQO,TORIC} but our geometric construct will not allow for these to be independent of one another.
In a related vein, the Kitaev Toric Code model symmetries associated with different windings of the torus can be continuously deformed. For the compass models that we detail in Section \ref{sec:POC}, any deformation of the original contours along which the products defining the vertical or horizontal symmetries will no longer remain a symmetry. 

\subsection{Exact universal exponential degeneracy of hybrids of thermal systems and their environment}

We conclude this Section
with two exceptionally simple yet general results relating to maxim {\bf{(B)}}. 

We consider a Hamiltonian $H_{\sf open}$ that is a sum of bonds $\{b_{\gamma}\}$ (Eq. (\ref{eq:bond:def})) with each of these bonds being invariant under all of the symmetries $\{U_{a}\}$
and $\{V_{a}\}$. For this Hamiltonian, we may establish 

{\bf{Corollary 4.}}
If a degeneracy that is an integer multiple of an exponential in the linear system length (or any other) is established for each level of $H_{\sf open}$ satisfying the requirements of Theorem 2 
then, a degeneracy that is an integer multiple of $2^{\cal{M}}$ also follows for the expectation value (as evaluated in eigenstates of $H_{\sf open}$) in the {\it same} Hilbert space, of any Hamiltonian $H_{\sf sub}$ that is formed by summing any subsets of the bonds appearing in $H_{\sf open}$.

{\it Proof.} The sequence of steps used to establish Theorem 2 can be repeated verbatim here since each of the individual bonds $b_{\gamma}$, whose sum forms $H_{\sf sub}$, commutes with all the symmetries $\{U_{a}\}$ and $\{V_{a}\}$. Thus, for each eigenstate $\ket{\psi}$ of $H_{\sf open}$, the expectation value $\langle\psi| H_{\sf sub}|\psi\rangle$ will remain invariant under the change of sign of any of the symmetry eigenvalues $\{\lambda_{1} \ldots \lambda_{a} \ldots \lambda_{{\cal{M}}}\}$. Since there are $2^{\cal{M}}$ such subsets, the corollary follows.   ~~$\blacksquare$

The subsets of the above bonds appearing
in $H_{\sf open}$ may tessellate any subregion
$\Omega_{\sf sub}$
of the domain $\Omega_{D}$. The above proof is rather general and also holds when subregion  $\Omega_{\sf sub}$ is not be geometrically contiguous. In what briefly follows, we consider what occurs when $\Omega_{\sf sub}$ is not composed of disjoint volumes. Formally, in the thermodynamic limit, we may regard $\Omega_{\sf sub} \subset \Omega_{D}$
as the ``system bulk'' and the remaining  $\Omega_{\sf env.} \equiv (\Omega_{D} - \Omega_{\sf sub})$ as the
thermal reservoir or ``environment'' with which it interacts and is in equilibrium with. This is so since the combined ``system-environment'' hybrid is described by $H_{\sf open}$. If this hybrid is in thermal equilibrium then its state will be a thermal state defined by $H_{\sf open}$. From Theorem 2 and Corollary 4, in each of the exponentially many degenerate ground states (which may be regarded as thermal states at temperature $T=0$) of the environment-system hybrid of $H_{\sf open}$, the energy of the system alone $\langle H_{\sf sub} \rangle$ is exactly the same. Identical results trivially hold for all excited states. Thus, the logarithm of the number of states of this hybrid having the same system energy $\langle\psi| H_{\sf sub}|\psi\rangle$ is bounded from below by ${\cal{O}}(L^{d'})$. The linear dimension $L$ of the system is less than or equal to than that of the ``system-environment'' hybrid $\Omega_{D}$ for which Theorem 2 applies. We stress that this conclusion holds for all possible Hamiltonians $H_{\sf open}$ satisfying the requirements of Theorem 2 that include the system $H_{\sf sub}$ as a subset. 
For any such Hamiltonian, we may deform the volume $\Omega_{\sf env.}$ defining the ``environment'' and make it arbitrarily small as long as the requirements of Theorem 2 are still satisfied. In particular, since $\Omega_{\sf sub} \subset \Omega_{D}$, for all such deformations of the environment that geometrically surrounds the system, we will consistently obtain a universal minimal lower bound of ${\cal{O}}(L^{d'})$ on the ground state entropy. This hints that the introduction of various boundary effects
in the thermodynamic limit may still leave the ground state entropy bounded by ${\cal{O}}(L^{d'})$. 

Although obvious, we should perhaps emphasize that, in general,
 $[H_{\sf open}, H_{\sf sub}] \neq 0$. Thus despite the fact that they may share the same eigenvalues of their common $d$-dimensional symmetry operators when diagonalized separately, $H_{\sf open}$ and $H_{\sf sub}$ do not have
identical eigenstates. That is, this corollary is not a trivial consequence of enlarging the Hilbert space of the system (on which $H_{\sf sub}$ operates) by the additional degrees of freedom of the environment. 

We may further trivially invert the roles of the ``system" and ``environment" to obtain an additional result. Towards this end, we may let the closed ``system-environment'' hybrid be a system for which we cannot prove by Theorem 2 the exponential degeneracy (i.e., when the conditions of this theorem are not met) and take the ``system" to be a theory on a volume $\Omega_{\sf sub}$ for which the conditions of Theorem 2 are satisfied. When ensemble equivalence holds (as expected in the thermodynamic $L \to \infty$ limit), this will lead to another very simple consequence.

{\bf{Corollary 5.}} 
Consider a subvolume ${\Omega_{\sf sub}}$ of linear dimension $L$ for which the conditions of Theorem 2 apply. We further assume an ensemble equivalence holds between the system properties on $\Omega_{\sf sub}$ and the system properties within the system-environment hybrid on $\Omega_{D}$, the reduced density matrix formed by a partial trace of the density matrix on the general volume $\Omega_{D}$ (for which the conditions of Theorem 2 need not apply). Under these conditions, the system ground state entropy on $\Omega_{\sf sub}$ will be $\ge {\cal{O}}(L^{d'})$. 

{\it{Proof.}} The von-Neumann entropy of the ``system" on $\Omega_{\sf sub}$ at any given temperature (including the limiting $T=0$ case associated with the ground state sector) is
\begin{eqnarray}
\label{sys-entropy}
S_{\sf sub} = - {\sf Tr} \rho_{\sf sub} \log_2 \rho_{\sf sub}.
\end{eqnarray}
Up to a factor of $k_{B} \ln 2$, this is the standard thermodynamic entropy of the system. In Eq. (\ref{sys-entropy}),
the reduced density matrix
\begin{eqnarray}
\label{rho-sys1}
\rho_{\sf sub} = {\sf Tr}_{\sf env.}  \rho_{\sf sub-env.} 
\end{eqnarray}
is formed by a partial trace of the density matrix of the system-environment $\rho_{\sf sub-env.}$ over the Hilbert space (${\sf env.}$) of the environment. In the microcanonical ensemble, the entropy of the system is given by the logarithm of the number of its ground states. Thus, from Theorem 2,
\begin{eqnarray}
\label{sys-s-bound}
S_{\sf sub} \ge {\cal{O}}(L^{d'}).
\end{eqnarray}
We next invoke ensemble equivalence. This equivalence implies that calculations with the reduced density matrix of Eq. (\ref{rho-sys1}) reproduce computations with the density matrix of the closed system (i.e., those within the microcanonical ensemble). Considering the ground state sector of the system-environment hybrid, we see from Eqs. (\ref{sys-entropy},\ref{sys-s-bound}), that whenever the system satisfies the requirements of Theorem 2, the partial trace of  Eq. (\ref{rho-sys1}) yields a reduced density matrix for the system with an entropy that is bounded from below by ${\cal{O}}(L^{d'})$. ~ $\blacksquare$

Similar entropy bounds may be derived on various system-environment hybrids that contain one or more subregions. All such bounds are consistent with known entropy inequalities \cite{AK}. However, albeit consistent with our bounds, the entropy inequalities of \cite{AK} do not rigorously demonstrate that the ground state degeneracy on {\it any} theory with higher symmetries on an arbitrary volume $\Omega_{D}$ (i.e., including those for which the conditions of Theorem 2 are not satisfied) must be exponentially large. Returning to the proof of Corollary 5, We emphasize that even if the ground states of the system-environment hybrid are not highly degenerate, for all ``system'' subvolumes $\Omega_{\sf sub}$ for which the conditions of Theorem 2 are met, the ground state entropy bounds of Eq. (\ref{sys-s-bound})
are satisfied. Notwithstanding special exceptions, e.g., \cite{julienb}, ensemble equivalence is quite pervasive in the thermodynamic $L \to \infty$ limit of rather general theories. In the above, we considered general ensembles formed by system-environment hybrids. In the conventional ``canonical" setting, the environment is far larger than the system while in the microcanonical ensemble, the system on $\Omega_{\sf sub}$ is closed and there is no environment. Here, ensemble equivalence in the form of the irrelevance of the fine details of the boundary environment with which the system interacts and equilibrates with is conceptually similar to independence of the system behavior from its boundary conditions. 

In order to make the content of Corollaries 4 and 5 clear, we will discuss their implications for the Square lattice compass model at the end of Section \ref{sec:cyl}. 

In Section \ref{sec:boundary1}, we will return to a general discussion of maxim {\bf{(B)}} as it pertains to the effects boundary conditions.

\section{Examples: Square and Cubic Lattice  Compass Models}
\label{sec:POC}

\subsection{Definition of the 90$^\circ$ square lattice model}
\label{POC:sq}

\begin{figure}[htb]
	\centering
	\includegraphics[scale=0.35]{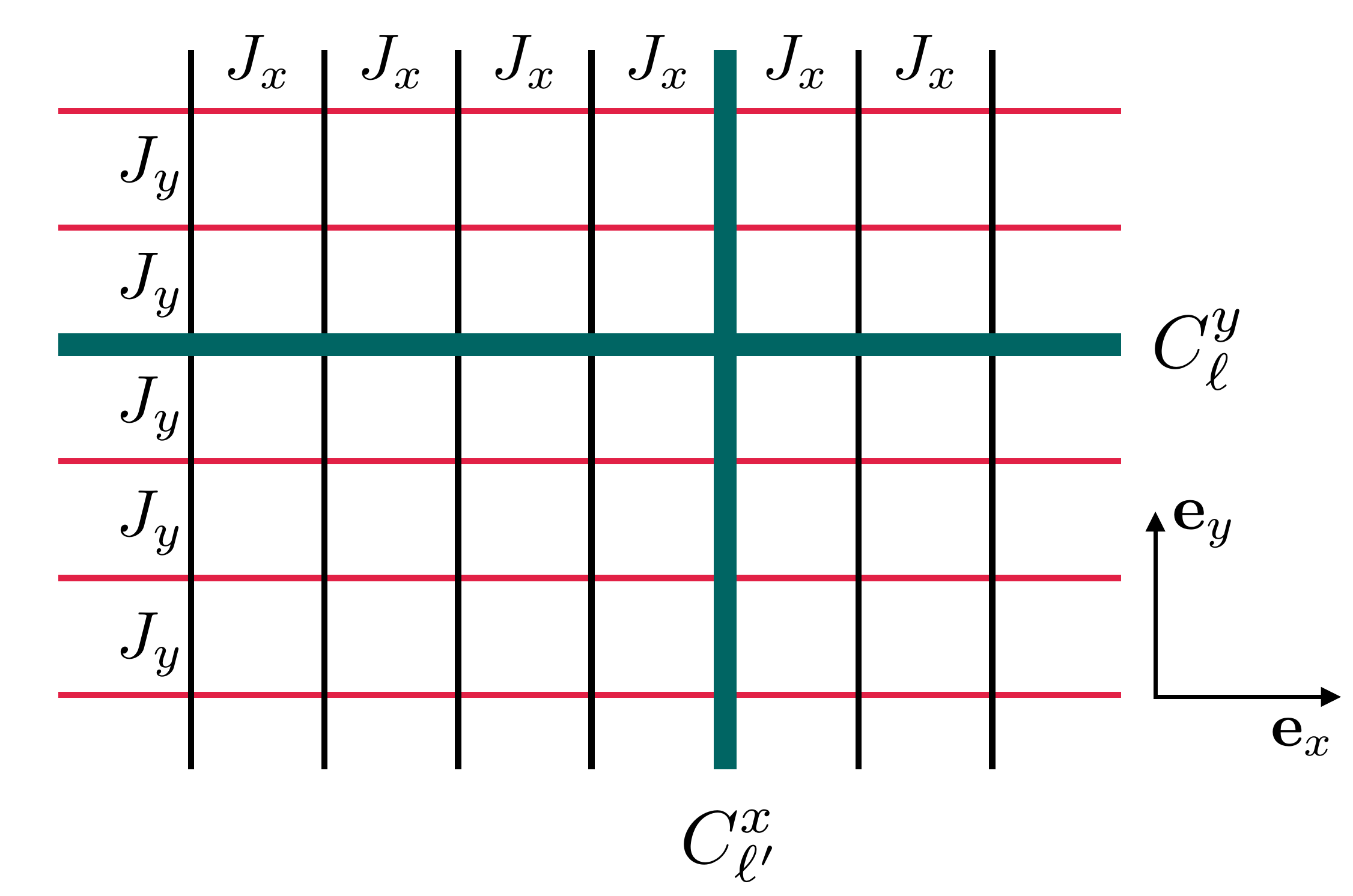}
	\caption{(Color online.)  The PCM, Eq. (\ref{ocmeq}), on an open square lattice. Interactions along the two external Cartesian directions are associated with couplings $J_\mu$, $\mu=x,y$. Corresponding with each of the highlighted contours $C_{\ell}^{x}$ and $C_{\ell'}^{y}$ are the symmetries of Eq. (\ref{sym2d}).}
	\label{fig:symm}
\end{figure}

As an example in which Theorem 2 comes to life, we will first consider the spin $S=1/2$ square lattice $90^\circ$ compass model, the simplest of all compass models \cite{NBCB,BCN,BN,Bacon,Ma,NF,DBM,NO,NOC1,rev-compass,Long-TQO,PNAS,PRLduality,AIP,KK,jan1,120compass,simon}. We start with a brief definition of this model. The 
90$^\circ$ square lattice or planar compass model (PCM) \cite{NF} on a square lattice of size $N= L \times L$
(with a lattice constant that we set to be unity) is given by
a simple bilinear in the Pauli operators, 
\begin{eqnarray} 
H_{\sf PCM} = -  \sum_{r,\mu=x,y} J_{\mu} \, \sigma^{\mu}_{r} \sigma^{\mu}_{r+
{\bf e}_\mu}.
\label{ocmeq}
\end{eqnarray}
Interactions (or ``bonds'') involving the $x$ spin operators $S^x_r= 
\sigma^x_r/2$ (here and throughout we set $\hbar=1$) occur only along the spatial $x$ direction of the lattice. 
Similar spatial direction-dependent spin exchange interactions
appear for the $y$ components of the spin. Thus, the PCM  displays exchange interactions 
with two Pauli matrix flavors along the $x$ and $y$ square lattice directions, 
which we henceforth refer to as ${\bf e}_\mu$ with $\mu =x,y$ (see Fig. \ref{fig:symm}). 
This model may be regarded as a square lattice rendition of Kitaev's honeycomb model \cite{Kitaev06} (in which all three different Pauli matrices appear along the three honeycomb lattice directions connecting nearest neighbors). 

Obviously, models exactly dual to the PCM will share identical spectra \cite{PRLduality,AIP}. For instance, 
the simple duality 
($J_{x} \leftrightarrow J_{\sf P}$ and $J_{y} \leftrightarrow h$) \cite{NF,AIP} maps the PCM of Eq. (\ref{ocmeq}) onto
a model first introduced by Xu and Moore \cite{XM},
\begin{eqnarray}
H_{XM} = - J_{\sf P} \sum_{\Box} \prod_{r \in \Box} \sigma^{z}_{r} - h \sum_{r} \sigma^{x}_{r}.
\label{eq:XM}
\end{eqnarray}
Here, $\prod_{r \in \Box} \sigma^{z}_{r}$ and $\sum_{\Box}$ denote, respectively, the product of the four spins lying at the vertices of minimal square lattice plaquettes $\Box$ and the sum over all such plaquettes $\Box$ in the square lattice.

\subsection{$d=1$ Gauge Like Symmetries of the square lattice $90^\circ$ compass model}
\label{sec:d=1}

The PCM is invariant under the following symmetries \cite{NF,benoit,DBM,BN,NO}
\begin{eqnarray}
{\hat{O}}^{\mu}_{\ell}=  \prod_{r \in C^{\mu}_{\ell}} i\sigma_{r}^{\mu} \ \mbox{ , for }
\mu = x,y
\label{sym2d}
\end{eqnarray}
with $C^{\mu}_{\ell} \perp {\bf e}_\mu$ axis (see Fig. \ref{fig:symm}). We briefly review some of the properties of these symmetries.
Our notation will follow that of \cite{NF,BN,NO} while the results pertaining to the commutation and anti-commutation 
of these symmetry operators amongst themselves were first reported in \cite{benoit,DBM}. 
All of the symmetries associated with lines $C^{\mu}_{\ell}$, that are all parallel to the same direction $\mu$, commute amongst themselves,
\begin{eqnarray}
[{\hat{O}}^{\mu}_{\ell} , {\hat{O}}^{\mu}_{\ell'}]=0.
\label{ooc}
\end{eqnarray}
By contrast, operators of the type of Eq. (\ref{sym2d}) that are related to orthogonal lines anticommute with one another,
\begin{eqnarray}
\label{ooa}
\{{\hat{O}}^{x}_{\ell}, {\hat{O}}^{y}_{\ell'}\}=0.
\end{eqnarray} 

The above anticommutativity is apparent since any two orthogonal lines $\ell$ and $\ell'$ intersect and share one common lattice site $r'$. Associated with this point of intersection, ${\hat{O}}^{x}_{\ell}$ has a single factor of $\sigma^{x}_{r'}$ in the string product of Eq. (\ref{sym2d}). Similarly, ${\hat{O}}^{y}_{\ell'}$ has, in the string product of Eq. (\ref{sym2d}) that defines it, a factor of $\sigma^{y}_{r'}$. The anticommutator $\{\sigma^{x}_{r'}, \sigma^{y}_{r'} \}=0$ implies the anticommutation
relation of Eq. (\ref{ooa}). For general boundary conditions discussed in the next subsections \cite{explain-square-compass-relation}, different independent composites of the two dual sets of $d=1$ symmetries  $\{{\hat{O}}^{y}_{\ell'}\}$
and 
$\{{\hat{O}}^{x}_{\ell} \}$ 
on horizontal and vertical lines
correspondingly relate, in the general setting of our central Theorem 2, to the two respective sets of operators $\{U_{a}\}$ and $\{V_{a'}\}$. \\

When $J_{x} = J_{y}$, a global reflection symmetry augments the symmetries of Eq. (\ref{sym2d}) \cite{Long-TQO}. We will, however, focus on the anisotropic model of Eq. (\ref{ocmeq}) with $J_{x} \neq J_{y}$. As $[H_{\sf PCM},\hat{O}^{\mu}_{\ell}]=0$, similar to Eq. (\ref{el-state}), we can label, with some abuse of notation, the eigenstates of the Hamiltonian $H_{\sf PCM}$ by 
\begin{eqnarray}
\label{e-state}
| \psi \rangle = |\lambda_{\mu;1} ... \lambda_{\mu;L}, \{\nu\} \rangle,
\end{eqnarray}
with $\lambda_{\mu;\ell} = \pm 1$ an eigenvalue associated with the mutually commuting symmetries ${\hat{O}}^{\mu}_{\ell}$ (all operators with the same value of $\mu$)
and $\{ \nu \}$ an additional label for all states within a given sector of $\{\lambda_{\mu;\ell}\}_{\ell=1}^{L}$. 
This latter label may mark the energies of
these states and, when additional degeneracies appear, any other remaining quantum numbers. 

In the classical ($S\rightarrow \infty$) limit, the PCM trivially exhibits a degeneracy that is exponential in the linear system length $L$. In this limit, the model transforms into that of classical two-component (XY) spins $\vec{S}_r=(S^x_r,S^y_r)$. Classically, the symmetries ${\hat{O}}^{x}_{\ell}$ and ${\hat{O}}^{y}_{\ell'}$ correspond, respectively, to a reflection of all of the $L$ spins $\vec{S}_r$ on a given vertical/horizontal line $\ell$ about the internal vertical/horizontal spin direction (i.e., $S^y_r \to S^y_r$ and $S^x_r \to -S^x_r$ for a reflection about the internal $S^y$ axis). Any spin $\vec{S}_r$ cannot be parallel to both the internal vertical and horizontal spin directions and thus will change under the application of (at least) one of the two symmetries $\hat{O}^{x}_{\ell}$ and $\hat{O}^{y}_{\ell'}$ associated with the vertical and horizontal lines that pass through the site $r$. One may traverse the real space square lattice along one of its diagonals and sequentially consider the $L$ sites $r_{\sf diag}$ along it. In doing so, one sees that at each site on the diagonal an application of (at least) one of the two symmetries  ${\hat{O}}^{x}_{\ell}$ and ${\hat{O}}^{y}_{\ell'}$ that have the lines $\ell$ and $\ell'$ intersect at $r_{\sf diag}$ will change the spin configuration. Thus, by turning ``on'' or ``off'' these $L$ reflection symmetries, one may generate $2^L$ states (i.e., XY spin configurations) that are degenerate with any given classical spin state. Complementing these exact symmetries, the classical system further exhibits an emergent continuous rotational symmetry that appears in its ground state sector. For instance, for positive coupling constants $J_{x}=J_{y}=J$, it is indeed
readily established that
all uniform (i.e., ferromagnetic) states are ground states of the Hamiltonian with continuous global rotation connecting these ground states. These global emergent symmetries remain unchanged in the classical limit for states formed by the application of any combination of the $d=1$ symmetries on the classical ferromagnetic states. Superficially similar to anomalies in quantum field theories, this emergent continuous symmetry of the classical system is no longer a symmetry of the quantum theory. 

In what follows, we focus our attention on the spin $S=1/2$ quantum model of Eq. (\ref{ocmeq}). 

\subsection{Open square lattice with boundaries parallel to the Cartesian directions}
\label{rev}

We first briefly summarize the standard open boundary condition realization of the $S=1/2$ PCM. We will
then turn to consider other boundary conditions and illustrate how they imply
an exponential degeneracy of the spectrum. Our bound on the degeneracy applies for {\it each level of the spectrum, not only the ground state sector}. 

When the square lattice is aligned along the Cartesian $x$ and $y$ directions, given any eigenstate
of the form of Eq. (\ref{e-state}) with $\mu$ fixed to be either $x$ or $y$, we can apply a symmetry operator
of the type ${\hat{O}}^{\mu' \neq \mu}_{\ell'}$ to $\ket{\psi}$. This leads to a new eigenstate of $H_{\sf PCM}$,
\begin{eqnarray}
\label{global-}
{\hat{O}}^{\mu' \neq \mu}_{\ell'} |  \lambda_{\mu;1} ... \lambda_{\mu;L}, \{\nu\} \rangle &=& 
 | -  \lambda_{\mu;1} ...-  \lambda_{\mu;L},\{ \nu \} \rangle\nonumber
 \\  &\equiv& | \psi' \rangle
 \end{eqnarray}
 that has the same energy
 as the original state $|\psi \rangle$. 
 The result of Eq. (\ref{global-}) is the same for all lines $C^{\mu'}_{\ell'}$ 
 with $1 \le \ell' \le L$.
 Thus, given any initial eigenstate $| \psi \rangle$, we see that
 we can construct a second eigenstate that has the
 same energy. A two-fold degeneracy \cite{DBM} thus follows from the existence of
 the non-commuting symmetries of Eqs. (\ref{sym2d}, \ref{ooa}). An identical
 effect and conclusion trivially follow from time-reversal symmetry (and Kramers degeneracy)
 as applied to square lattices with an odd $L \times L $ size \cite{NO}.
 On a square lattice having its edges parallel to the Cartesian directions, no additional
 degeneracy follows from symmetries. Indeed, numerically only a two-fold degeneracy is observed on finite size square lattices \cite{benoit,DBM}. Indeed, as our proofs make clear (requiring the need for the conditions underlying Theorem 2 to be satisfied), {\it the existence of higher symmetries does not imply an exponentially large degeneracy of all finite size systems}. A curious numerical observation \cite{DBM} of the PCM on finite size square lattice (insofar as finite size calculations can suggest) 
 is that as the system size becomes progressively larger, sets of $2^{L}$ states each seem to become degenerate as $L \to \infty$. The results of our work will further rationalize these findings. We first explain why in open systems with various boundaries, the symmetries of Eq. (\ref{sym2d}) mandate a degeneracy of each level
 that is exponential in the perimeter. 
 
\begin{figure}[htb]
	\centering
	\includegraphics[scale=0.25]{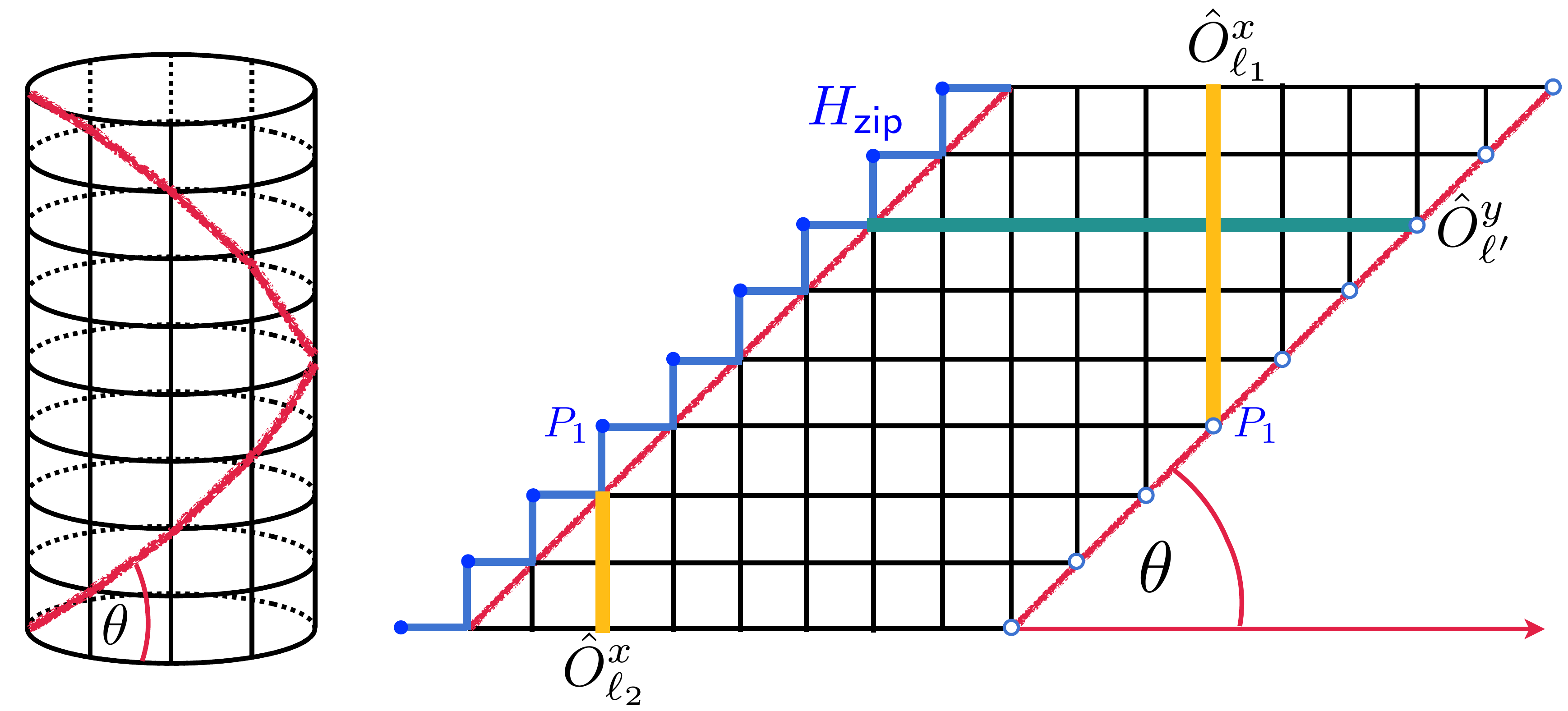}
	\caption{(Color online.)  The system of Eq. (\ref{ocmeq}) on a square ($L=8$) lattice with cylindrical boundary conditions (black grid, left panel) and its cut along a slanted line (highlighted in red). The cut leads to the system shown on the right. On the original cylinder, the two (yellow) lines marked by $P_1$ were one and the same. The nearest-neighbor PCM Hamiltonian  on the open parallelogram on the right can be made to agree with that on the cylinder by inserting additional interactions (staircase blue line) that become non-local on the parallelogram. The sum of these additional $2L-1$ interaction terms defines the ``zipper Hamiltonian'' $H_{\sf zip}$. Both the cylindrical system on the left and the open parallelogram on the right are invariant under $d=1$ symmetries of the type of Eq. (\ref{sym2d}). Both, open and cylindrical, systems are invariant under the symmetries ${\hat{O}}^{y}_{\ell'}$ associated with the horizontal lines $C^y_{\ell'}$ (marked in green). The open system is invariant under the symmetries 
	${\hat{O}}^{x}_{\ell}$ affiliated with all vertical lines $C_{\ell}^{x}$ connecting one side of the open system to the other. However, since the vertical boundaries of the cylindrical system differ from those of the open system, its symmetries differ. For instance, individually, while the two  operators ${\hat{O}}^{x}_{\ell_2=3}$ and  ${\hat{O}}^{x}_{\ell_1=11}$  (marked in yellow) are exact symmetries of the open system Hamiltonian $H_{\sf open}$ they are no longer symmetries of the cylindrical system. However, the product ${\hat{O}}^{x}_{3} {\hat{O}}^{x}_{11}$ is that of Eq. (\ref{sym2d}) associated with spanning a vertical generator of the cylinder and, as such, is a symmetry of the cylindrical system.}
	\label{fig:cylinder}
\end{figure}

 \subsection{``Cylindrical cuts'', ``toroidal cuts'', ``dilute vacancies'' and other boundary conditions or changes of internal geometry for which exponential degeneracy follows from symmetries}
 \label{sec:cyl}

 We next consider, see Fig. \ref{fig:cylinder}, a finite $L \times L$ square lattice which lies inside a parallelogram of an angle $\theta$ relative to the $x$ axis (with the $x$ and $y$ coordinates defining the nearest-neighbor compass interactions of Eq. (\ref{ocmeq})). This lattice can be viewed as that generated by ``cutting'' and ``opening up'' an $L \times L$ square lattice with cylindrical boundary conditions. If the cut were performed along a Cartesian direction ($\theta=\pi/2$ in Fig. \ref{fig:cylinder}), the result would be the standard square lattice with open boundary conditions. If, however, the cut is performed at another general angle $\theta$, the resulting parallelogram will define the lattice that we focus on now. This parallelogram will be composed of $L$ horizontal lines (each of length $L$). In Fig. \ref{fig:cylinder}, we show a parallelogram formed by choosing $\theta = \pi/4$.
 In traversing the lattice by a distance of order unity along the vertical direction, these lines are horizontally displaced relative to one another by one lattice constant. Thus, overall, the horizontal span of the lattice is equal to $(2L)$ lattice constants.
We will denote the Hamiltonian of Eq. (\ref{ocmeq}) on the so-generated open cylindrical cut by $H_{\sf open}$. 
  As earlier, the operators of Eq. (\ref{sym2d})
are symmetries for lines $C^{\mu}_{\ell}$ orthogonal to the $\mu$ direction. 
Now, however, we have the lines $ \{C^{x}_{\ell}\}_{\ell=1}^{2L}$ and
$\{C^{y}_{\ell'}\}_{\ell'=1}^{L}$. Clearly, if two eigenstates have a differing set of $\lambda_{y;\ell'}$
eigenvalues then they will be orthogonal to one another,
\begin{eqnarray}
\label{ortho} \hspace*{-0.7cm}
\langle  \lambda_{y;1} ... \lambda_{y;L}, \{\nu\} |  \lambda'_{y;1} ... \lambda'_{y;L}, \{\nu\}  \rangle 
= \prod_{\ell'=1}^{L} \delta_{\lambda^{\;}_{y;\ell'}, \lambda'_{y;\ell'}}.
\end{eqnarray}
From the symmetry operators of Eq. (\ref{sym2d}) we can construct the subset of composite symmetry
operators given by
\begin{eqnarray}
\label{composite}
{\hat{O}}^{x}_{\sf {\cal{S}}} = 
\prod_{\ell \in {\cal{S}}} {\hat{O}}^{x}_{\ell},
\end{eqnarray}
where ${\cal{S}}$ is any of the possible subsets of the integers $\{1,2,3, ..., L\}$ labeling the $L$ leftmost vertical lines.
Note that here we allow for the action of the independent vertical symmetry operators $\{{\hat{O}}^{x}_{\ell}\}$ only on the $L$ leftmost vertical lines out of the $(2L)$ vertical lines. The effect of higher $\ell>L$ vertical symmetry operators ${\hat{O}}^{x}_{\ell}$ on the eigenvalues $\lambda_{y;1} ... \lambda_{y;L}$ can be expressed in terms of that of lower $\ell$ operators $\{{\hat{O}}^{x}_{\ell}\}_{\ell=1}^{L}$. That is, with reference to
Fig. \ref{fig:cylinder}, for all $1 \le \ell < L$, the result of applying ${\hat{O}}^{x}_{\ell} {\hat{O}}^{x}_{L+ \ell}$ on the set of eigenvalues $\lambda_{y;1} ... \lambda_{y;L}$  is the same as that of applying
${\hat{O}}^{x}_{L}$ (which simply flips the sign of all of these eigenvalues). There is an exponentially large number (i.e., $\sum_{k=0}^{L} \binom{L}{k} = 2^{L}$) of such independent subsets of products amongst the vertical symmetry operators. On the square lattice of subsection \ref{rev},
the application of these symmetry operators in this subset can lead to one of two outcomes: if 
the set ${\cal{S}}$ contains an even number of integers then given any eigenstate of
the form of Eq. (\ref{e-state}), the product ${\hat{O}}^{x}_{\sf {\cal{S}}} | \psi \rangle$ will give back the same full set of symmetry eigenvalues $\lambda_{y; \ell'}$.
By contrast, if the set ${\cal{S}}$ contains an odd number of integers $1 \le \ell \le L$ then ${\hat{O}}^{x}_{\sf {\cal{S}}} | \psi \rangle = | \psi' \rangle$ with $| \psi' \rangle$ being the state defined in Eq. (\ref{global-}). Now, here is one of trivial yet nonetheless crucial points that we wish to bring to the fore:
The application of the symmetry operations of Eqs. (\ref{composite}) on 
an initial state $\ket{\psi}$ will give rise to an exponential number of orthogonal states,
\begin{eqnarray}
\label{oys}
{\hat{O}}^{x}_{ \sf {\cal{S}}} | \psi \rangle = |    \lambda^{\cal{S}}_{y;1} ...  \lambda^{\cal{S}}_{y;L}, \{\nu\}  \rangle \equiv | \psi^{\cal{S}} \rangle.
\end{eqnarray}
Here, $\lambda^{\cal{S}}_{y;\ell'} = \eta_{\ell'} {\lambda}_{y;\ell'}$. For a given $\ell'$, the Ising variable $\eta_{\ell'} = \pm 1$ denotes the even/odd parity of the number of vertical lines $C^{x}_{\ell}$ associated with the set ${\cal{S}}$ (i.e.,
$\ell \in {\cal{S}}$) that intercept the line $C^{y}_{\ell'}$. Each of the possible different general choices for the set ${\cal{S}}$ will uniquely lead to a different binary string ${\cal{B}} = (\lambda_{y;1}^{\cal{S}} ... \lambda_{y;L}^{\cal{S}})$.
By the orthogonality relation of Eq. (\ref{ortho}), this implies an exponentially large number of degenerate orthogonal eigenstates. In particular, since each amongst the possible different choices of the strings 
$(\lambda_{y;1} = \pm 1,  \ldots, \lambda_{y;L}= \pm 1)$ can be achieved then the system will have a degeneracy which is (at least) of size $2^{L}$. More generally, the degeneracy
as associated with the $\lambda_{y;\ell}$ eigenvalues alone is bounded by the cardinality ${\cal{M}}= |{\cal{B}}| = | {\cal{S}}|$ of the set of attainable binary strings by applying the different operators $\{{\hat{O}}^{x}_{\sf {\cal{S}}}\}$ as in Eq. (\ref{oys}).
For general $\theta$, the number of such binary strings scales as $|{\cal{B}}| = 2^{L}$. When $\theta=\pi/2$, the number of different obtainable binary strings (and thus a lower bound
on the degeneracy) is $|{\cal{B}}|=2$ (as in subsection \ref{rev}). If, in approaching the thermodynamic limit, the same degeneracy is found 
irrespective of the tilt angle $\theta$ then we see how a degeneracy of $2^{L}$ must appear for a square lattice oriented along the Cartesian axes. 

In fact, this is an example of the consequences of our general Theorem 2. Towards this end, we may identify $U_1 = {\hat{O}}^y_{\ell'=1}$, $V_1 = {\hat{O}}^x_{\ell=1}$,
 $U_{2}={\hat{O}}^y_{\ell'=2}$, 
 $V_2 = ({\hat{O}}^x_{\ell=1}{\hat{O}}^x_{\ell=2})$, $\ldots, U_{1<a\le L} = {\hat{O}}^y_{\ell'=a}$, $V_{1<a\le L} = ({\hat{O}}^x_{\ell=a-1} {\hat{O}}^x_{\ell=a}), ~\ldots$ so that ${\cal{M}} = 2^L$. (See \cite{explain-square-compass-relation} for further comparison to the case of conventional boundary conditions.)

Interestingly, our open parallelogram construct is a bipartite lattice with sublattices $\Lambda_A$ and $\Lambda_B$ of equal cardinality. One can construct the unitary operator
\begin{eqnarray}
\label{chiral:eq}
{\cal U}_{\sf ch}=\prod_{r \in \Lambda_A} \sigma^z_r ,
\end{eqnarray}
such that $\{H_{\sf open},{\cal U}_{\sf ch}\}=0$, meaning that the spectrum of $H_{\sf open}$ is symmetric with respect to zero and, thus, ${\cal U}_{\sf ch}$ is a chiral symmetry. This result is valid for any bipartite lattice.

Thus far, in this subsection, we considered a cut of the cylinder of Fig. \ref{fig:cylinder} to generate the open system of $H_{\sf open}$
for which we can establish the exponential degeneracy. Along nearly identical lines, we may similarly examine a torus that is cut a general angle $\theta \neq \pi/2$ and opened up to as produce an oblique cylinder. For such general angles $\theta$, Theorem 2 will then imply the existence of an exponential ground state degeneracy on the resultant oblique cylinders. 

Similar constructs may be devised for many other boundary geometries. For instance, placing the square lattice on a closed cone (with the generating line of the cone defining the $y$-axis and its base parallel to the $x$-axis) instead of the cut cylinder of Fig. \ref{fig:cylinder}, would, for general opening angles of the cone, enable a proof a degeneracy that would scale, once again, exponentially in the linear system dimension. One may similarly examine a hybrid of two such cones sharing the same base and in an analogous fashion also for a trapezoid. A
particular realization of these other geometries is that of the ``Aztec diamond'' lattice (a square lattice rotated at 45$^{\circ}$ so as to have a diamond shaped boundary) \cite{Aztec1,Aztec2,Aztec3}.

More generally, a repetition of the above steps for generic geometries in which the top and bottom boundaries (or, similarly, left and right boundaries) are shifted relative to one another by a distance that is ${\cal{O}}(L)$, will yield a lower bound of the logarithm of the degeneracy that is of order ${\cal{O}}(L)$. 

Further yet, numerous other open boundaries exist other than the above boundary conditions that correspond to different cuts of cylindrical or toroidal systems. As the reader can indeed see, the boundaries of generic open two-dimensional systems of linear size $L$ will allow for ${\cal{O}}(L)$ independent $d=1$ symmetries. Thus, {\it the logarithm of degeneracy of general physical realizations of two-dimensional systems will scale as ${\cal{O}}(L)$}. The same considerations can be extended to arbitrary dimensions.

In addition to modifying boundary conditions to go to the open parallelogram or other systems for which we can prove our theorem, we can also insert modify the  {\it internal geometry} by inserting {\it dilute vacancies} such that the conditions of Theorem 2 will be satisfied. For the PCM, we can remove a single site from every column at a different height. 
For the PCM on an $L \times L$ square lattice, if we insert $L$ vacancies such that they satisfy the above condition of one vacancy per each column at different heights, then an exponential degeneracy is assured by Theorem 2.

Before concluding this subsection, we briefly return to Corollaries 4 and 5. 

In the context of corollary 4, we note that if the region  $\Omega_{D}$ is a parallelogram (as it is
in the current example of the PCM) then we may take its subset to, e.g., be a square that can be inscribed in the parallelogram $\Omega_{D}$. For situations in which we can prove exponential degeneracies for $H_{\sf open}$ augmented by a number of additional boundary terms (e.g., any number of horizontal bonds in our example of the PCM and a reduction by a factor of a half in our lower bound on the degeneracy when any vertical bond was added), we may similarly remove any subset of the bonds to establish, at least, the same degeneracies.

With reference to corollary 5, we may take $\Omega_{D}$ to be of any shape (including the open square lattice of Section \ref{rev}) and $\Omega_{\sf sub} \subset \Omega_{D}$ to be a parallelogram of linear scale $L$ inscribed within it.  $\Omega_{\sf sub}$ may similarly be any other subvolume of $\Omega_{D}$ such that the Hamiltonian $H_{\sf open}$ on it satisfies the requirements of Theorem 2. Corollary 5 then asserts that if ensemble equivalence holds (as expected in the thermodynamic limit), given the $T=0$ ground state sector density matrix on $\Omega_{D}$, the reduced density matrix of Eq. (\ref{rho-sys1}) on $\Omega_{\sf sub}$ yields an entropy satisfying Eq. (\ref{sys-s-bound}). 

\subsection{The cubic lattice $90^\circ$ compass model}

\begin{figure}[htb]
	\centering
	\includegraphics[scale=0.45]{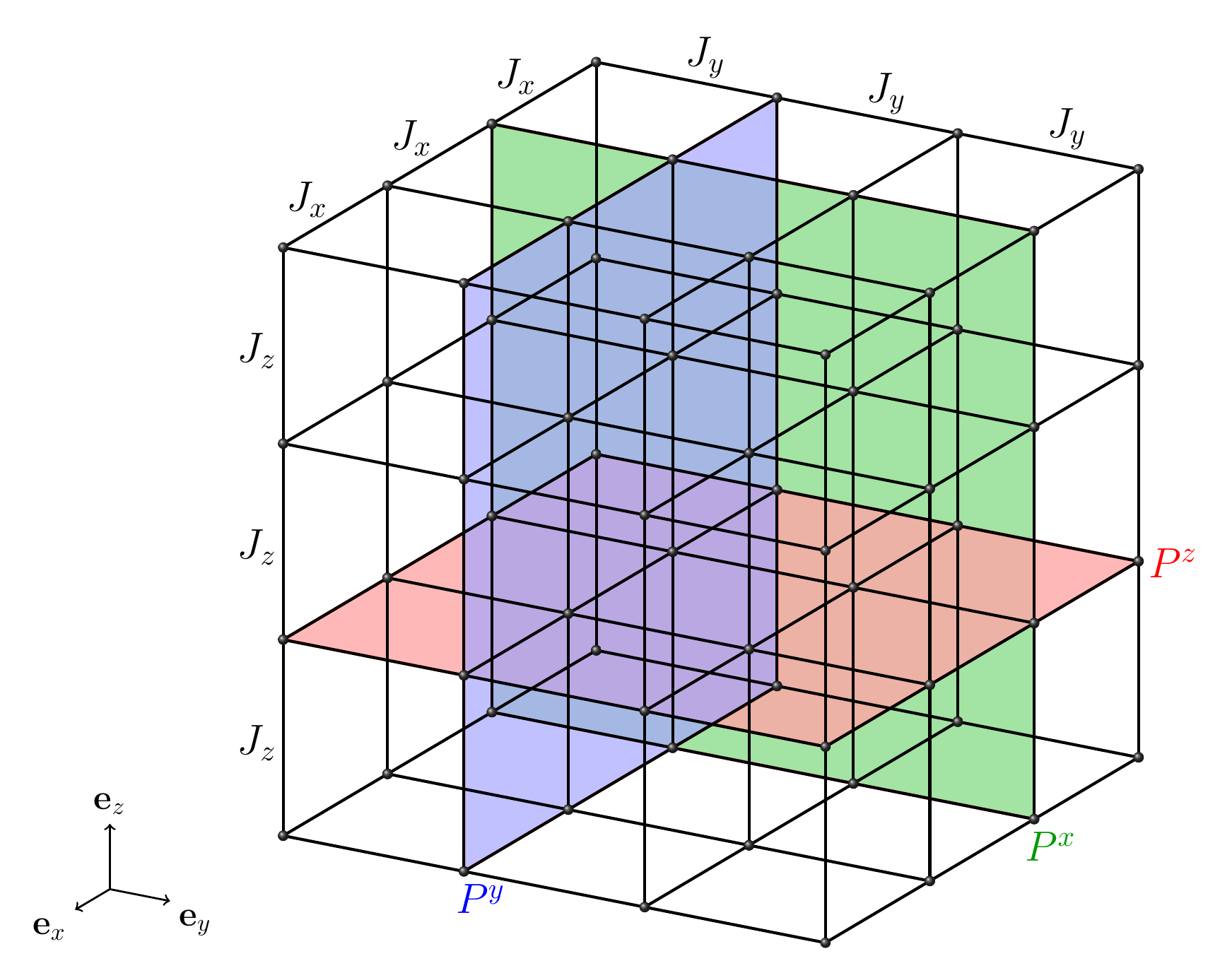}
	\caption{(Color online.) The cubic lattice compass model and its symmetries of Eq. (\ref{sym3d}) (see text). }
	\label{fig:cubic}
\end{figure}

Complementing Kitaev's honeycomb model \cite{Kitaev06}, the above square lattice compass model may be trivially extended to other geometries \cite{rev-compass}. In particular, the cubic ($ L \times L \times L $) lattice $90^\circ$ compass model Hamiltonian $H_{\sf CCM}$ is given by the righthand side of Eq. (\ref{ocmeq}) with the sum over $\mu$ now spanning the three external spatial Cartesian directions $\mu=x,y,z$ and the three associated internal spin components (see Fig. \ref{fig:cubic}). This model features $d=2$ symmetries given by Eq. (\ref{sym2d}) where,
on the cubic lattice, $\mu = x,y,$ and $z$ and $C^{\mu}_{\ell} \perp {\bf e}_\mu$ now become $d=2$ dimensional planes $P\perp {\bf e}_\mu$. That is, rather explicitly,
\begin{eqnarray}
{\hat{O}}^{\mu}_{P}=  \prod_{r \in P} i\sigma_{r}^{\mu} \ \mbox{ , for }
\mu = x,y,z  \ .
\label{sym3d}
\end{eqnarray}

Similar to the discussion in Section \ref{sec:d=1},
in the large spin classical ($S \to \infty$) limit, the $d=2$ symmetries that this system hosts become reflections about internal spin directions. The exponential degeneracy in the number of planes $L$ similarly follows. We now discuss the $S=1/2$ quantum model.

With the trivial replacement of the $d=1$ dimensional line label $\ell$ by that of the $d=2$ dimensional plane $P$, the symmetries ${\hat{O}}^{\mu}_{P}$ satisfy the commutation relations of Eq. (\ref{ooc}), i.e., 
$[\hat{O}^{\mu}_{P},\hat{O}^{\mu}_{P'}] =0$.
Whenever there are, at the intersection of two orthogonal planes $P\perp {\bf e}_\mu$ and $P' \perp {\bf{e}}_{\mu'\neq \mu}$, an odd number of sites, we will find the $D=3$ dimensional analog of 
Eq. (\ref{ooa}), i.e., the anticommutator
\begin{eqnarray}
\label{3D-}
\{{\hat{O}}^{\mu}_{P}, {\hat{O}}^{\mu'}_{P'}\}=0.
\end{eqnarray} 
As in the square lattice compass model, the eigenstates of the Hamiltonian $H_{\sf CCM}$ will be of the form of Eq. (\ref{e-state}), 
$| \psi \rangle = |\lambda_{\mu;1} ... \lambda_{\mu;L}, \{\nu\} \rangle$ with $\{\lambda_{\mu_{1}}, \lambda_{\mu_{2}}, \ldots \lambda_{\mu_{L}}\}$ now being the set of eigenvalues $\{\lambda_{\mu_{P}}\}$ of the symmetries  $\{{\hat{O}}^{\mu}_{P}\}$ associated with the $L$ parallel planes $\{P\}$ that are orthogonal to ${\bf e}_\mu$. So long as we ensure that intersecting orthogonal planes share an odd number of common sites so that Eq. (\ref{3D-}) is satisfied, all of the previous considerations that detailed above for the square lattice may be repeated for the cubic lattice. In particular, if we consider boundary conditions satisfying this intersection property then we may examine three-dimensional variants of the deformed boundary conditions of Section \ref{sec:cyl}. Given the $L$ symmetry eigenvalues $\{\lambda_{\mu_{P}}\}$, repeating the considerations for the square lattice compass model for the cubic compass model, we find that {\it each energy eigenvalue} (whether that of a ground state or an arbitrary excited state) enjoys, once again, a degeneracy that is an integer multiple of $2^L$.

\subsection{A $U(1)$ symmetric cubic compass model}
\label{sec:rotsym}

All of the symmetries in the compass model examples discussed thus far were of a discrete ($\mathbb{Z}_2$) nature. A simple model harboring a continuous $U(1)$ symmetry is given by the cubic lattice Hamiltonian 
\begin{eqnarray}
H_{\sf rot.~symm.} = -J_{x} \sum_{r} (\sigma^{y}_{r} \sigma^{y}_{r+ {\bf e}_{x}} + \sigma^{z}_{r} \sigma^{z}_{r+ {\bf e}_{x}}) \nonumber
\\ -J_{y} \sum_{r} (\sigma^{x}_{r} \sigma^{x}_{r+ {\bf e}_{y}} + \sigma^{z}_{r} \sigma^{z}_{r+ {\bf e}_{y}}) \nonumber
\\ -J_{z} \sum_{r} (\sigma^{x}_{r} \sigma^{x}_{r+ {\bf e}_{z}} + \sigma^{y}_{r} \sigma^{y}_{r+ {\bf e}_{z}}).
\label{eq:rotsym}
\end{eqnarray}
This Hamiltonian is invariant under continuous rotations in the planes $P\perp {\bf e}_\mu$,
\begin{eqnarray}
\label{rotplane}
\prod_{r \in C^{\mu}_{P}} e^{i \theta \sigma_{r}^{\mu}/2 }.
\end{eqnarray}
In the taxonomy of Refs. \cite{BN,NOC1,NO,Long-TQO,PNAS,PRLduality,AIP,rev-compass},
the operators of Eq. (\ref{rotplane}) are 
$d=2$ dimensional (since the planes $P$ are two-dimensional) $U(1)$ symmetries. On an $L \times L \times L$ lattice, there are $3L$ such planes $P$ and thus $3L$ symmetries of the type of Eq. (\ref{rotplane}). From a generalization of Elitzur's theorem \cite{BN,NOC1}, continuous $d=2$ symmetries in systems having a spectral gap cannot be broken even at zero temperature nor can they be broken at finite temperatures (i.e., positive energy density above the ground state) in both gapped and gapless systems. In the context of the focus of the current paper, it follows that if gapped renditions of the above system, two-dimensional (and/or other continuous) $U(1)$ symmetries cannot be broken. 

In what follows, when referring to examples, the canonical model that we will refer to will largely be that of the spin $S=1/2$ PCM of the earlier subsections. The $U(1)$ symmetric model introduced in this subsection illustrates how continuous higher symmetries may appear in lattice systems.

\section{Effects of boundary conditions on spectral degeneracies}
\label{sec:boundary1}

We next turn to the influence of the boundaries on the degeneracies. We will do so by amending the open system Hamiltonian $H_{\sf open}$ on the volume $\Omega_{D}$ by additional terms (whose sum is the ``zipper Hamiltonian'' $H_{\sf zip}$) that will capture the effects of changing the boundary conditions. In effect, we will ``stitch'' back the open system to form a closed surface. Intuitively, one may anticipate $H_{\sf zip}$ to not radically influence the spectrum of the bulk system. As we will argue, the effect of $H_{\sf zip}$ on the system degeneracy may indeed be rather modest (vanishing in the thermodynamic limit) in certain situations.
Augmenting Corollaries 4 and 5, this will strengthen the plausibility of maxim {\bf (B)} of the Introduction as it pertains to boundary effects. To make our discussion clear, we first explain what $H_{\sf zip}$ is for the planar compass model of Eq. (\ref{ocmeq}) and then outline results for general systems.

We proceed with the ``surgery'' outlined in Fig. \ref{fig:cylinder}. Following the incision of the lattice on the cylinder (left panel of that figure) and the ``flattening'' of the square lattice leading to a open parallelogram boundary (right panel), we may ``sew'' back the missing cut links associated with the lattice sites that lie on the boundaries of the open square lattice. We do so as to exactly reproduce the Hamiltonian of Eq. (\ref{ocmeq}) on the original square lattice on the cylinder. 

Within the coordinate frame of the open square lattice, all of the interactions contained in $H_{\sf zip}$ (linking one boundary site on one side of the system to another on the opposite side) are long ranged (of spatial separation that is of the order of the system size ${\cal{O}}(L) \to \infty$ in the thermodynamic limit). Referring to Fig. \ref{fig:cylinder}, we may write 
\begin{eqnarray}
\label{zip-define}
\! \! \! \! \! \! H_{\sf zip} &\equiv&  -  \sum_{r \in B_{L}}  
(J_{x} \, \sigma^{x}_{r} \sigma^{x}_{r+ (L-1)
{\bf{e}}_{x}} \nonumber
+ J_{y} \, \sigma^{y}_{r} \sigma^{y}_{r+(L-1){\bf{e}}_{x} + {\bf{e}}_{y}})  \nonumber \\
&\equiv& \  \sum_{\partial \gamma}   b_{\partial \gamma}.
\label{zipper}
\end{eqnarray}
The unitary operator ${\cal U}_{\sf ch}$ of Eq. (\ref{chiral:eq}) anticommutes with $H_{\sf zip}$ implying that like $H_{\sf open}$ it, too, trivially has a spectrum that is symmetric about zero. For the PCM, the index $\partial \gamma$ in Eq. (\ref{zipper}) labeling the boundary bonds forming $H_{\sf zip}$ is comprised of the spatial location $r \in B_{L}$ marking the line of sites lying to the left of the ``cylindrical'' cut and the index $\mu=x,y$ denoting the internal components of the spins appearing in the first equality of Eq. (\ref{zipper}).  In the (uncut) cylindrical system on the left panel of Fig. \ref{fig:cylinder}, these long range terms $b_{\partial \gamma}$ correspond to nearest neighbor interactions (see also the staircase on the righthand panel of that figure), and  $H_{\sf zip}$
is a one dimensional Hamiltonian having a rather trivial spectrum (Appendix \ref{sec:e-zip}). The complication arises for the full spectral problem posed by the cylindrical Hamiltonian 
\begin{equation}
\label{coz}
H_{\sf cyl} = H_{\sf open}+ H_{\sf zip}  .
\end{equation}

In the zipper Hamiltonian bonds associated with interactions between spins that lie on the same row commute with all of the symmetries ${\hat{O}}^{y}_{\ell'}$ and thus do not lift the degeneracy of the states
of Eq. (\ref{composite}) associated with their eigenvalues. By contrast, the vertical bonds in $H_{\sf zip}$ do not commute with these symmetries and thus may lift the earlier found exponential degeneracy.

Indeed, both the open cylindrical cut Hamiltonian $H_{\sf open}$ and the closed cylinder Hamiltonian $H_{\sf cyl}$ commute with the horizontal line symmetries $\{{\hat{O}}^{\mu}_{\ell'}\}$ of Eq. (\ref{sym2d}) (and thus, trivially, with their products). However, it is only for $H_{\sf open}$ that we are able to rigorously establish the existence of an exponential degeneracy for arbitrary linear system size $L$. In the PCM, if we add (i.e., ``stitch back'') a subset of $L_{\sf zip}^{\sf vert}$ vertical bonds 
that appear in Eq. (\ref{zipper}) and further similarly add (or ``sew back'') all of the horizontal bonds 
appearing in Eq. (\ref{zipper}) to $H_{\sf open}$ then we will be left with  $(L-L_{\sf zip}^{\sf vert})$ horizontal symmetries $\{{\hat{O}}^{\mu}_{\ell'}\}$. Repeating {\it mutatis mutandis} the steps of Section \ref{sec:POC}, we will then be able to prove that each of the eigenstates has a degeneracy that is an integer multiple of $2^{L-L_{\sf zip}^{\sf vert}}$.

In the common eigenbasis of the symmetries $U_{a}$ (i.e., the symmetries ${\hat{O}}^{y}_{\ell'}$ commuting with $H_{\sf cyl}$) and $H_{\sf cyl}$, the effect of ${\hat{O}}^{x}_{\ell}$ is to flip the sign of the single vertical line $C^x_\ell$ that has $P$ as one of its endpoints. Furthermore, the open string ${\hat{O}}^{x}_{\ell}$ flips the sign of all of the ${\hat{O}}^{y}_{\ell'}$ eigenvalues  $\{\lambda_{1}, \ldots, \lambda_{L}\}$ associated with horizontal lines $C^y_{\ell'}$ that intersect with $C^x_\ell$. In the ground state, one would expect all of the vertical links to have a negative contribution to the energy. 

Although the first equality of Eq. (\ref{zipper}) is specific to the PCM, the more abstract second equality
may apply to any system.
In all such systems,
when summed over all $\partial \gamma$, the individual long range boundary bond operators will
form the relevant zipper Hamiltonian. In general,
the boundary bonds $b_{\partial \gamma}$ are functions of operators 
(fields) $\phi_{r}^{\mu}$ associated
with the boundary $r \in B_{L}$. 
Whenever we change the boundary conditions for an arbitrary local Hamiltonian (that may differ from that of the PCM), we will be able to
express $H_{\sf zip}$ as a sum of such local terms. Similarly, instead of sewing the boundaries of the open parallelogram to form a cylinder, one may repopulate the {\it dilute vacancies} discussed towards the end of Section \ref{sec:cyl}  to form other zipper type Hamiltonians (in which the Hilbert space will now be increased) that are a sum of local terms.

Generically, the full Hamiltonian involving $H_{\sf zip}$ does not commute with $H_{\sf open}$. This non-commutativity makes the analysis using symmetries less obvious. Nonetheless, rather broad conclusions may still be drawn. In what follows, we  outline the general structure of the exact eigenstates when the effects of the zipper Hamiltonian $H_{\sf zip}$ are included. We then discuss the diagonal matrix elements of $H_{\sf zip}$ in the eigenbasis of $H_{\sf open}$ in systems with a spectral gap. The latter result shows that, to lowest order in perturbation theory, exponential degeneracy remains asymptotically unchanged in the thermodynamic limit. Henceforth, our discussion will become more qualitative. 

In Appendix \ref{sec-high-heuristic}, we review and extend the discussion of perturbation theory of a different sort. This will motivate us, in Appendix \ref{varsec}, to introduce simple variational states that suggest trends that are consistent with those numerically observed in \cite{benoit,DBM}.

We now return to the general problem (not focusing our discussions to the example of the PCM) and reformulate some of our considerations more broadly. Since the symmetries $\{U_{a}\}$ are not lifted by $H_{\sf zip}$, it follows that the eigenstates
 of $H_{\sf cyl}$ may be expressed as a linear superposition of eigenstates of $H_{\sf open}$, in general of different energy 
 (Eq. (\ref{el-state})), with the same symmetry eigenvalues, i.e., as
\begin{eqnarray}
| \lambda_1 \lambda_2 \ldots \lambda_{L}, \{{\nu}_{\sf cyl} \}\rangle &=& \nonumber
\\ && \hspace*{-2.0cm}\sum_{\{\nu\}} c_{\lambda_1 \lambda_2 \ldots \lambda_{L}, \{\nu\}}
|\lambda_{1} \lambda_{2} \ldots \lambda_{\cal{M}}, \{\nu\} \rangle.
\label{eq:proj}
\end{eqnarray}
We underscore that in the above sum, sectors of differing quantum numbers $\{\nu\}$ of the open system eigenstates may be mixed. By contrast, different eigenvalues of the symmetry operators $U_a$ {\it cannot} be superposed. Thus, rather trivially, $H_{\sf zip}$ cannot directly lift the degeneracy associated with the symmetries $U_{a}$ by mixing states with different eigenvalues of these symmetries. In a related vein, since $H_{\sf cyl}$ commutes with the symmetries $\{U_{a} \}$, it follows that all of the projected Hamiltonians $P_{\{\nu\}}H_{\sf cyl} P_{\{{\nu'\}}}$ are block diagonal in the eigenspace of $H_{\sf open}$. This structure of Hamiltonian (and ensuing form of its eigenstates) is reminiscent of that in textbook type \cite{AM} Bloch systems wherein mixing may occur only between states, differing by reciprocal lattice vectors $\vec{K}$, in different Brillouin zones (playing the role of $\{\nu\}$) yet not between states belonging to the same Brillouin zone (with crystal momentum labels $\vec{k}$ in the first Brillouin zone replaced here by the eigenvalues $\lambda_1 \lambda_2 \ldots \lambda_{L}$).
Augmenting Eq. (\ref{eq:proj}), as the commutation relations $[H_{\sf cyl}, U_{a} ]=0$, for all $a=1,2, \ldots, {\cal{M}}$, further make clear, symmetry considerations trivially reduce the eigenvalue problem of $H_{\sf cyl}$ to that in the $2^{\cal M}$ decoupled sectors labelled by different eigenvalue strings $\{\lambda_1 \ldots \lambda_{\cal{M}}\}$. In each such sector, we need to diagonalize the projected Hamiltonian $P_{\lambda_1 \ldots \lambda_{\cal{M}}} H_{\sf cyl} P_{\lambda_1 \ldots \lambda_{\cal{M}}}$ with $P_{\lambda_1 \ldots \lambda_{\cal{M}}}$ 
denoting the projection operator to a set of fixed higher symmetry eigenvalues. The union of the eigenvalues of these projected Hamiltonians over all $\{\lambda_1 \ldots \lambda_{\cal{M}}\}$ trivially forms the complete spectrum of $H_{\sf cyl}$. Since $[H_{\sf open}, V_{a}]=0$, the eigenvalues of the projected Hamiltonians $P_{\lambda_1 \ldots \lambda_{\cal{M}}} H_{\sf open} P_{\lambda_1 \cdots \lambda_{\cal{M}}}$
are the same in different sectors  $\{\lambda_{a}\}_{a=1}^{\cal{M}}$- this is, in essence, the content of Lemma 1.
As we noted previously, for the very same reasons, the demonstration of exponential degeneracy can be further extended to the eigenstates of $H_{\sf cyl}^{c} \equiv H_{\sf open} + H^{c}_{\sf zip}$ where $H^{c}_{\sf zip}$ is the sum of all of the bonds appearing in $H_{\sf zip}$ that commute with all of the operators $\{V_{a}\}_{a=1}^{\cal{M}}$. In the context of the PCM, $H^{c}_{\sf zip}$ is the sum of all horizontal boundary terms appearing in Fig. \ref{fig:cylinder} connecting sites that lie along opposite sides of the cylindrical cut. However, as we further underscored for the full the zipper Hamiltonian $H_{\sf zip}$ (not the sum $H^{c}_{\sf zip}$ of the subset of bonds in $H_{\sf zip}$ that commute with $\{V_{a}\}_{a=1}^{\cal{M}}$), due to Eq. (\ref{coz}) the commutator $[H_{\sf zip}, V_{a}] = [H_{\sf cyl}, V_{a}] \neq 0$ for general $V_{a}$. As a consequence of this non-vanishing commutator, the eigenvalue problem of $P_{\lambda_1 \ldots \lambda_{\cal{M}}} H_{\sf cyl} P_{\lambda_1 \ldots \lambda_{\cal{M}}}$ in different sectors $\lambda_{1} \ldots \lambda_{\cal{M}}$ is not identically the same. This discrepancy led to the aforementioned possible removal of the spectral degeneracy. 

As we elaborate in Appendix \ref{A:symm}, the Wigner-Eckart theorem may be naturally extended for higher symmetries. This trivial generalization leads to selection rules on non-vanishing matrix elements of general operators (including the zipper Hamiltonian) in the common eigenbasis of $H_{\sf open}$ and its symmetries.

The expectation values of $H_{\sf zip}$ in the cylindrical states $| \lambda_1 \lambda_2 \ldots \lambda_{L}, \{{\nu}_{\sf cyl} \}\rangle$ of Eq. (\ref{eq:proj}) are trivially equal to the energies associated with $H_{\sf zip}$ on these states on the closed cylinder. Since $H_{\sf zip}$ is a sum of short range terms on the cylinder, the expectation values in these cylindrical states are, generally, finite (also within the thermodynamic $L \to \infty$ limit). For the translationally invariant ground states on the cylinder, these expectation value will equal to the average expectation per bond within the global ground state of $H_{\sf cyl}$ times the number of bonds appearing in $H_{\sf zip}$. 

We now briefly turn to a somewhat more specific discussion of theories (dependent on local operators $\{\phi^{\mu}_{r}\}$) that exhibit higher symmetries. In the eigenbasis of $H_{\sf open}$, the diagonal matrix elements of $H_{\sf zip}$ are the expectation values of the sum of these interactions ($H_{\sf zip} = \sum_{\partial \gamma} b_{\partial \gamma}(\{\phi^{\mu}_{r}\})$) in the eigenstates of $H_{\sf open}$. Note that in any eigenstate $\langle \psi| b_{\partial \gamma}| \psi \rangle$ is a sum of correlation functions involving the boundary local operators $\{\phi^{\mu}_{r}\}$. 
If the connected correlation function between local observables in eigenstates of $H_{\sf open}$ decays exponentially in distance between these observables then we may state another simple Lemma. \\

{\bf Lemma 6.}
Consider the open system Hamiltonian $H_{\sf open}$ to be a sum of {\it interactions of finite range} and strength (Eq. (\ref{eq:bond:def})) such that the eigenstates of $H_{\sf open}$ do not support, in any eigenstate, infinite range nor algebraic correlations between local boundary fields $\phi^{\mu}_r$ (i.e., for asymptotically large separation distance $L$, the correlation functions are bounded by $C + {\cal{O}}(e^{-L/\xi})$ with $C$ being a constant that depends only on the energy eigenvalue of $H_{\sf open}$, and $\xi$ a characteristic correlation length). Under these conditions, the diagonal elements of $H_{\sf zip}$ in the eigenbasis of $H_{\sf open}$ (i.e., $\langle \psi|H_{\sf zip} | \psi \rangle = \sum_{\partial \gamma} \langle \psi| b_{\partial \gamma} (\{\phi_{r}^{\mu}\})| \psi \rangle$) with $|\psi \rangle$ an eigenstate of $H_{\sf open}$) must tend to a uniform constant in the $L \to \infty$ limit \cite{explain_SPT}. 

{\it Proof.}
$H_{\sf zip}$ is a sum of, at most, ${\cal{O}}(L^{D-1})$ boundary interactions between local boundary operators $\phi^{\mu}_{r}$ that are a distance ${\cal{O}}(L)$ apart. Given the assumption above, each of the interaction terms $b_{\partial \gamma}$ formed by their products has, in a sector of fixed eigenvalue of $H_{\sf open}$, a constant expectation value of $C$ up to exponentially small corrections in $L$. This implies that the diagonal matrix elements of $H_{\sf zip}$ are a sum of ${\cal{O}}(L^{D-1})$ individual expectation values $\langle b_{\partial \gamma} \rangle$
that, up to the above stated uniform shift, are each bounded by decaying exponential in $L$. Such a sum is bounded from above by a number of order ${\cal{O}}(L^{D-1} e^{-L/\xi})$ and thus tends to zero in the $L \to \infty$ limit.
   ~$\blacksquare$

Thus, in the thermodynamic limit, the expectation values $\langle H_{\sf cyl} \rangle$ in each of the exponentially many degenerate eigenstates of $H_{\sf open}$ are the same. We reiterate that, as our proof of Lemma 6 illustrates, this property {\it emerges only in the asymptotic $L \to \infty$ limit}. For finite $L$, there are additional deviations ${\cal{O}}(L^{D-1} e^{-L/\xi})$
about the asymptotic uniform constant value of $\langle H_{\sf cyl} \rangle$. 

In those models in which $H_{\sf zip}$ is a positive semidefinite operator and $C=0$ in the ground state manifold of $H_{\sf open}$, there is an (asymptotic) exponentially large degeneracy in the ground state of $H_{\sf cyl}$. The proof of the latter assertion is nearly immediate as we now explain. In such all such models, each of the ground states of $H_{\sf open}$ of which, by Theorem 2, there are exponentially many, may be taken as variational ground states for $H_{\sf cyl}$. In each of these (linearly independent) variational states \cite{MacDonald, JonesOrtiz}, Lemma 6 asserts that the expectation value of $H_{\sf zip}$ tends, in the thermodynamic limit, to zero.
On the other hand, positive semidefinite Hamiltonians $H_{\sf zip}$ imply that the ground states of $H_{\sf open}$ provide lower bounds to $H_{\sf cyl}$. It follows that $H_{\sf cyl}$ enjoys precisely the same (asymptotic) exponentially large (in $L^{d'}$) ground state degeneracy in the thermodynamic limit. 

In Appendix \ref{expdecay}, we will review general arguments for the exponential decay of correlations in the ground states of Lorentz invariant gapped systems that admit a Wick rotation. If the thermal averages may be replaced by eigenstate averages (in the spirit of the Eigenstate Thermalization Hypothesis \cite{ETH1,ETH2,ETH3,ETH4}) then it follows that all diagonal matrix elements of $H_{\sf zip}$ vanish in the thermodynamic limit. 

In Appendix \ref{lopt}, we demonstrate that
Lemma 6 mandates that to lowest order in degenerate perturbation theory (in the perturbation $H_{\sf zip}$ that alters the system boundary conditions), the exponential degeneracy of $H_{\sf open}$
is not lifted.

\section{Theories with UV/IR mixing displaying conventional IR behavior}

We now discuss the physical consequences of the degeneracies that we rigorously established for various boundary conditions and suggested for others. As we reviewed in the Introduction, the application of higher symmetries naturally leads to a mixing of IR and UV modes. This mixing is apparent in the ground states. Already at the classical level, the low energy field configurations involve both Fourier modes of both very low and very high wavenumbers. 
Similar to ($d=0$) gauge symmetries, also $d=1$ discrete symmetries or $d=2$ continuous symmetries that give rise to low energy short wavelength variations of the fields {\it cannot}, by the generalized Elitzur theorem \cite{BN,NOC1}, be spontaneously broken. Just as in gauge theories in which the local symmetries cannot be spontaneously broken, these low $d-$dimensional Gauge Like symmetries do not preclude the existence of usual thermodynamic transitions. Systems with higher $d\ge 2$ discrete symmetries (that may be spontaneously broken) can exhibit rather conventional behaviors. In some systems, including those with symmetries that are not exact but rather only emerge at low energies as symmetries of the projected ground state susbspace (e.g., the classical 120$^{\circ}$ compass model that exhibits emergent discrete $d=2$ symmetries in its ground state sector \cite{NBCB,BCN,rev-compass}), entropic fluctuations may stabilize ordering about a uniform state \cite{BCN,NBCB,rev-compass}. Where rigorous results exist, it is seen that at positive temperatures, whenever spontaneous symmetry can occur, UV/IR mixing may be lifted and the system will display long wavelength fluctuations about the low energy, higher symmetry allowed \cite{BN,NOC1}, orders \cite{BCN,NBCB,rev-compass}. Such finite temperature entropic ``order by disorder'' stabilization effects   appear in numerous other systems \cite{rev-compass,OBD,OBD1,OBD2,OBD3}.
As we alluded to in the Introduction, such symmetries may lead to a proliferation of minimizing modes on $d-$ dimensional (``flat band'' type) surfaces in $k$ space \cite{rev-compass}. Such numerous low energy states can also lead to glassy dynamics and rich spatial structures \cite{NAG, JoergPete,competing}. As we further noted in the Introduction, in several theories featuring higher symmetries, an exact dimensional reduction may occur as may be established via dualities, e.g., \cite{rev-compass,NOC1,zack,zack1}. The associated dual lower dimensional models \cite{rev-compass,NOC1,zack,zack1}
may belong to universality classes with conventional IR behaviors. These dualities have been used to establish exponential degeneracies in other systems, such as those trivially encountered when mapping spins to Majorana fermions in high dimensional interacting (Hubbard type and other) systems, e.g.,  \cite{Fermionization1}.  

\section{Conclusions}
The central result of this paper is that of Theorem 2. Systems displaying higher symmetries that are embedded on general open geometries display an exponentially large degeneracy. The situation for the conventional textbook Born-von Karman (i.e., periodic) boundary conditions is more subtle. We discussed conditions under which this degeneracy may persist when the geometry is deformed to be that of the Born-von Karman or other types by the addition of external environments with which the system interacts or the insertion of a ``zipper Hamiltonian."

\bigskip\noindent
{\bf Acknowledgements:} 
We are extremely grateful to Lauren Pusey-Nazzaro and Masaki Oshikawa for numerous discussions, critical reading of the work, and comments. We are also grateful for discussions with Alexander Abanov, Shmuel Nussinov, and Gilles Tarjus and to a question posed by Kirill Shtengel. Part of this work was performed at the Aspen Center for Physics (where some of our results were also presented) which is supported by National Science Foundation grant PHY-1607611. G.O. acknowledges support from the US Department of Energy grant DE-SC0020343.  

\appendix

\onecolumngrid

\section{The eigenvalues of the compass model Hamiltonian $H_{\sf zip}$}
\label{sec:e-zip}

Much of our discussion focused on the projected $H_{\sf zip}$ in the the common eigenbasis of $H_{\sf open}$ and the symmetries in a sector fixed $ \{\nu\}$ and $ \{\nu'\}$. We may also readily compute the eigenvalues of $H_{\sf zip}$ sans such a projection. Towards that end, we simply view $H_{\sf zip}$ as a one-dimensional Hamiltonian on the chain formed by the ``zipper''. This system may be trivially diagonalized by performing a Jordan-Wigner transformation followed by a Bogoliubov transformation. This system was indeed investigated since it may be regarded as a one-dimensional variant \cite{oles} of the Kitaev honeycomb model \cite{Kitaev06}. When $J_x=J_y=J$,
we can express, up to an innocuous additive constant, $H_{\sf zip}$ as
\begin{eqnarray}
\label{JW}
H_{\sf zip} = \sum_{k} \epsilon_{k}(c_{k}^{\dagger} c_{k} - \frac{1}{2}), \nonumber
\\ \epsilon_{k} = 2J(1- \cos k),
\end{eqnarray}
with $c_{k}$ and $c_{k}^{\dagger}$ the (spinless) Fermi annihilation and creation operators and $k$ a momentum label along the zigzagging zipper path (of total length $(2L-1)$), i.e., $k = \pi n/L$ with $n=1-L, \ldots L$. Since, for each $k$, the eigenvalues of the Fermi number operator $c_{k}^{\dagger} c_{k}$ are $0$ and $1$, the full eigenvalues of $H_{\sf zip}$, similar to the eigenvalues of its submatrices discussed above, are also sums or differences of individual terms (in this case, $\epsilon_{k}/2$). Each of these eigenvalues is $2^{N-2L+1}$ degenerate if we examine $H_{\sf zip}$ on the original Hilbert space of the $N$ spins lying on the cylinder of Fig. \ref{fig:cylinder}.

\section{Wigner-Eckart type selection rules in systems with higher symmetry and sparsity of the zipper Hamiltonians}
\label{A:symm}

Independent bounds, fortifying the considerations underlying our results, arise from the Wigner-Eckart theorem \cite{Eckart,Wigner1959,Wigner1941} as in its common applications to SU(2) and other symmetry groups \cite{Wigner-Eckart_theorem-General}. Here, we extend these to the matrix elements of the few body operators $b_{\partial \gamma}$ in the higher symmetry operator eigenbasis \cite{Long-TQO}. The arising symmetry constraints will, in particular, demand that the each of the operators $b_{\partial \gamma}$ is a sparse matrix when written in the eigenbasis of $H_{\sf open}$ spanned by the states of Eq. (\ref{el-state}). To see why this is so, we first consider as a general illustrative example, the textbook situation of an SU(2) symmetry eigenbasis labelled by the eigenvalues of the total squared angular momentum ${\sf J}^2$ (eigenvalue ${\sf j(j+1)}$) and ${\sf J}_{z}$ (eigenvalue ${\sf m}$). Specifically, for the ${\cal M}$ independent higher form symmetries, we consider situations in which we express the zipper Hamiltonian $H_{\sf zip}$ as a sum 
of products of tensors $\{T^{j_{\ell} m_{\ell}}_{\ell}\}_{\ell=1}^{\cal M}$ that transform irreducibly under each of these higher symmetries. With the full eigenvalue spectrum of $H_{\sf open}$ labelled by any additional multiplet index $\{\nu\}$, the Wigner-Eckart theorem states that for any irreducible operator $T_{\ell}^{\sf k_{\ell}q_{\ell}}$,
\begin{eqnarray}
\label{WE:eqnq}
&& \Big\langle  {\sf j'}_{1} {\sf m'}_{1} \ldots {\sf j'}_{\ell} {\sf m'}_{\ell} \{\nu'\} \ldots  {\sf j}'_{\cal M} {\sf m}'_{\cal M}, \{ \nu'\} \Big| T_\ell^{\sf k_{\ell}q_{\ell}} \Big|  ~{\sf j}_1 {\sf m}_1 \ldots {\sf j}_{\ell} {\sf m}_{\ell} \ldots \ldots  {\sf j}_{\cal M} {\sf m}_{\cal M},\{\nu\} \Big\rangle \nonumber
\\ &&= 
\frac{1}{\sqrt{2{\sf j_{\ell}}+1}} \Big\langle  {\sf j'}_{1} \ldots {\sf j'}_{\ell}  \ldots  {\sf j}'_{\cal M}, \{\nu'\} ~ \Big| T_{\ell}^{\sf k_{\ell}}\Big|  ~{\sf j}_{1}  \ldots  {\sf j}_{\ell} \ldots {\sf j}_{\cal M}, \{\nu\}\Big\rangle
\Big\langle {\sf j}'_{\ell} {\sf m}'_{\ell}; {\sf k_{\ell}q_{\ell}}\Big| {\sf j}_{\ell} {\sf m}_{\ell} \Big\rangle,
\end{eqnarray}
with 
$\langle {\sf j}'_{\ell} {\sf m}'_{\ell}; {\sf k_{\ell}q_{\ell}}| {\sf j}_{\ell} {\sf m}_{\ell} \rangle$ denoting 
the Clebsch-Gordan coefficients associated with the $d-$dimensional Gauge Like symmetry of the $\ell$-th layer. Thus, for any Hamiltonian $H_{\sf zip}$ that spans, at most, the spatial support of $R$ independent $d-$dimensional symmetries disjoint regions, the matrix element between two orthogonal
eigenstates of the $d \ge 1$-dimensional symmetry operator
\begin{eqnarray}
\Big\langle  {\sf j'}_{1} {\sf m'}_{1} \ldots {\sf j'}_{\ell} {\sf m'}_{\ell}  \ldots  {\sf j}'_{\cal M} {\sf m}'_{\cal M}, \{\nu'\}~ \Big| T_\ell^{\sf k_{\ell}q_{\ell}} \Big| ~ {\sf j}_1 {\sf m}_1 \ldots {\sf j}_{\ell} {\sf m}_{\ell} \ldots \ldots  {\sf j}_{\cal M} {\sf m}_{\cal M}, \{\nu\} \Big\rangle =
 0
\label{same}
\end{eqnarray}
in numerous instances. Given its above definition, $R$ may be viewed as a measure of the ``range of the interaction'' 
If when contrasting the ket and bra of Eq. (\ref{same}), more than $R$ different eigenvalue pairs $({\sf jm})$ differ from one another then the associated matrix element of $ T_\ell^{\sf k_{\ell}q_{\ell}}$ will vanish. Furthermore, for those few symmetry eigenvalues that do differ from one another (i.e., those marking off-diagonal elements in the symmetry eigenbasis), additional constraints will appear if $T_\ell^{\sf k_{\ell}q_{\ell}}$ are products of a finite number of single body operators (each of finite maximal angular momentum). Thus, when the off-diagonal matrix elements are nonzero, since the total $(d-$dimensional) angular momentum that a local operator can carry is finite the eigenvalue differences $|{\sf j}_{\ell} - {\sf j'}_{\ell}|$ and $|{\sf m}_{\ell} - {\sf m'}_{\ell}|$ can assume, at most, system size independent values of order unity. Such constraints can indeed be readily extended for symmetries other than SU(2). Applying these symmetry selection rules illustrates that, for the compass and other local models, the matrix elements of various local operators in the eigenbasis of $H_{\sf open}$ of Eq. (\ref{el-state}) vanish. Indeed, $H_{\sf zip}$ is diagonal in the symmetry projected eigenbasis of $H_{\sf open}$. The application of the higher symmetry variant of the Wigner-Eckart theorem to other local operators generally illustrates that these may only have sparse non-vanishing matrix elements. 

\section{Exponential decay of the matrix elements
of the projected $H_{\sf zip}$ in the eigenbasis of general gapped systems as suggested by Wick rotations in Loretnz invariant theories}
\label{expdecay}

We next briefly review considerations for exponential decay of correlations in gapped systems and extensions thereof to off-diagonal matrix elements.
An insightful approach \cite{matth} for establishing the exponential decay of spatial correlations relies on quasi-adiabatic processes and the Lieb-Robinson bounds \cite{LR}. In what follows, we discuss and extend an earlier method that relates correlations and evolution in time to those in space. Specifically, we will briefly touch on an analytic continuation (specifically a Wick rotation) of Lorentz invariant theories or their effective non-relativistic limit.
This will allow us to relate temporal correlations (that are naturally associated with spectral gaps)
to spatial correlations.
We will extend the dynamics provided by a general Hamiltonian $H_{\sf open}$ with eigenvectors $\{ |m \rangle \}$ to the Euclidean domain and illustrate that when the respective energy difference of the associated eigenstates is finite then the diagonal matrix elements of the additional bonds ($H_{\sf zip}$) that appear when the boundary conditions are changed, 
\begin{eqnarray}
\label{zipzero}
\lim_{L \to \infty} \langle \lambda_{\mu;1} ... \lambda_{\mu;L}, \{\nu'\} | H_{\sf zip} | \lambda_{\mu;1} ... \lambda_{\mu;L}, \{\nu\} \rangle = 0. \end{eqnarray}

To establish Eq. (\ref{zipzero}), we trivially apply standard Euclidean space demonstrations of the decay of correlations in the presence of a spectral gap (i.e., that of the diagonal component of the field bilinears) as in, e.g., Ref. [\onlinecite{QCDineq}]. We will consider, for a general Hamiltonian, equal time matrix elements of the product of two fields (spins). We define 
\begin{eqnarray}
\label{Fmr}
F^{\mu}(r,r')  \equiv \langle 0| \phi^{\mu} _{r}  \phi^{\mu}_{r'}  | 0 \rangle = \sum_{m}  \langle 0| \phi^{\mu} _{r} |m \rangle \langle m| \phi^{\mu}_{r'}  | 0 \rangle
\end{eqnarray}
with $| 0 \rangle$ denoting a ground state and $|m \neq 0 \rangle$ an excited eigenstate of $H_{\sf open}$. We next focus on each of the matrix elements in the sum of Eq. (\ref{Fmr}) and evaluate these by analytic continuation to the field theory to Euclidean space where we can invoke rotational invariance. In the original theory with the two fields  $\phi^{\mu}_{r}$ and $\phi^{\mu}_{r'}$ lie on the same time slice and the separation $(r'-r)$ is, obviously, space-like. We can express the two fields in terms of displacements from the field at the origin, 
\begin{eqnarray}
 \phi^{\mu} _{r}  = e^{- i p r}~ \phi^{\mu}_0 ~e^{i pr}, \nonumber
 \\ \phi^{\mu} _{r'}  = e^{- i p r'}~ \phi^{\mu}_0~ e^{i pr'}.
 \end{eqnarray}
 On inserting the complete set of eigenstates of the Hamiltonian $H_{\sf open}$, Eq. (\ref{Fmr}) becomes
 \begin{eqnarray}
\! \! \! \! \! \! \! \! \! \! \! \! \!  F^{\mu}(r,r')  = \sum_{m}  \langle 0|  e^{- i p r}~ \phi^{\mu} _{0} ~e^{i pr} |m \rangle \langle m|  e^{- i p r'}~ \phi^{\mu} _{0} ~e^{i pr'}   | 0 \rangle. 
\end{eqnarray}
We next use the aforementioned translational and rotational invariance of the Euclidean theory to evaluate $F^{\mu}(r,r')$. With the aid of these symmetries, we may translate and rotate the spatial coordinates $r \to (0, \vec{0})$ and $r' = r'-r= \to (\pm |t'-t|,\vec{0})$ so they both lie along the time axis of the Euclidean theory; $|t'-t| = |r'-r|$. This yields $e^{- i p r'}~ \phi^{\mu} _{0} ~e^{i pr'}   \to e^{\mp H(t'-t)} ~ \phi^{\mu} _{0} ~ e^{\pm H(t'-t)}$ (with the sign in the exponential chosen so as to ensure a well-defined analytic continuation). This yields
\begin{eqnarray}
\label{Wick}
 F^{\mu}(r,r')  = \sum_{m}  \langle 0|  \phi^{\mu} _{0}  |m \rangle  \langle m|  \phi^{\mu} _{0}  |0 \rangle e^{-(E_{m}- E_{0})|r'-r|}. 
  \end{eqnarray}
  To obtain the connected correlation function, i.e., the contribution to $F^{\mu}(r,r')$ from the excited states, we subtract from 
  Eq. (\ref{Wick})
  the ground state products
  $ \langle 0|  \phi^{\mu} _{0}  |0 \rangle  \langle 0|  \phi^{\mu} _{0}  |0 \rangle$.
  As its exponential factor makes clear,  Eq. (\ref{Wick}) then suggests that, in gapped systems, the connected correlation function must decay exponentially in the spatial separation $|r-r'|$. 
  Eq. (\ref{Wick}) therefore illustrates that if $H_{\sf zip}$ contains ${\cal{O}}(L)$ terms that each connect sites separated by a distance $|r'-r| = {\cal{O}}(L)$ then, in the presence of finite gaps $(E_{m}- E_{0})$, the diagonal matrix elements of $H_{\sf zip}$ in the eigenbasis of $H_{\sf open}$ will vanish in the $L \to \infty$ limit. Although, one may expect individual off-diagonal matrix elements of bounded local operators to decay with increasing system size (as in, e.g., the Eigenstate Thermalization Hypothesis \cite{ETH1,ETH2,ETH3,ETH4}), the full contribution of the (exponential in size) number of off-diagonal matrix elements to the energy eigenvalues is more complex.
  Clearly not all off-diagonal matrix elements (i.e., those between different excited states $n$) can be uniformly bounded such their sum vanishes exponentially with the distance $|r-r'|$ when the square norm $(\phi_{0}^{\mu} \phi_{r}^{\mu})^{\dagger} (\phi_{0}^{\mu} \phi_{r}^{\mu}) $ is a constant and thus so is its expectation value in any state, e.g., $\langle n| (\phi_{0}^{\mu} \phi_{r}^{\mu})^{\dagger} (\phi_{0}^{\mu} \phi_{r}^{\mu}) | n \rangle = 
  \sum_{m}| \langle n|\phi_{0}^{\mu} \phi_{r}^{\mu} | m \rangle|^2 = {\sf const}$. 
  An example of a situation in which such a square norm is constant is that of the Pauli bilinears appearing in the zipper Hamiltonian of the compass model (Eq. (\ref{zip-define})). When $\phi_{r}^{\mu}$ are Pauli operators, the latter ``${\sf const}$'' is equal to one.

\section{Lowest order degenerate perturbation theory}
\label{lopt}
In this brief Appendix, we demonstrate 

{\bf Theorem 7.} In the thermodynamic limit, to lowest order in degenerate perturbation theory, the exponential degeneracy of $H_{\sf open}$ is not lifted by the perturbation $H_{\sf zip}$. 

{\it Proof.} To lowest order in degenerate perturbation theory, we need to diagonalize $H_{\sf zip}$ in the eigenbasis of $H_{\sf open}$. By Lemma 6, 
in {\it any} eigenstate $| \psi \rangle$ of $H_{\sf open}$ of the same fixed energy eigenvalue, the expectation value
\begin{eqnarray}
\label{vanishb}
\langle \psi| b_{\partial \gamma}(\{ \phi^{\mu}_{r}\}) | \psi \rangle = C + {\cal{O}}(e^{-L/\xi}).
\end{eqnarray}
Thus, in the thermodynamic $L \to \infty$ limit, the expectation value of any of the terms appearing in $H_{\sf zip}$ will tend to zero (or other uniform constant value). Since this assumption of Lemma 6 is made for {\it any} eigenstate of $H_{\sf open}$ \cite{explain_SPT}, 
it will, in particular, include also those specific eigenstates 
that diagonalize $H_{\sf zip}$. Because the latter eigenstates are the zeroth order eigenstates in degenerate perturbation theory having the eigenvalues as the first order corrections to the energy, we see that all corrections due to $H_{\sf zip}$ correspond to no change at all (or a uniform energy shift). Thus, 
{\it to first order in perturbation theory},
when $L \to \infty$,  {\it the $2^{\cal{M}}$ fold degeneracy of $H_{\sf open}$ is not lifted by $H_{\sf zip}$}.   ~$\blacksquare$

We briefly remark in \cite{compasscorr} on our PCM example where $H_{\sf zip}$ is a bilinear in the boundary spins and $\langle H_{\sf zip} \rangle$ becomes a sum of two-point correlators. In systems with a spectral gap, the ground state correlation functions decay exponentially with the distance between the local fields.

\section{Another (higher order) perturbation theory}
\label{sec-high-heuristic}

We now briefly discuss a different perturbative approach that generalizes those introduced in \cite{benoit,DBM} for the PCM. The perturbation theory that we consider is that for rather general systems that exhibit higher symmetries. Specifically, we consider systems in $D$ spatial dimensions displaying $d$-dimensional symmetries ($d<D$) for which the Hamiltonian can be expressed as
\begin{eqnarray}
\label{H0V}
H = H_{0} + H^{'}.
\end{eqnarray}
Here, $H_0$ is a Hamiltonian that has its support on ${\cal M}={\cal{O}}(L^{d'})$ decoupled $d-$dimensional regions $\{{\cal R}_{a}\}$ where the $d-$dimensional higher symmetries operate. The
perturbative ``interaction'' $H^{'}$ couples these $d-$dimensional regions to one another. In what follows, we will ask what transpires when $H^{'}$ is a sum of local operators. Following the steps that led to Lemma 1, in the absence of the interaction term $H{'}$, each level has a degeneracy that is an integer multiple of $2^{\cal M}$. The corresponding basis set of eigenstates can be expressed as the tensor product 
\begin{eqnarray}
\label{tensorp}
|  \lambda_1 \lambda_2 \ldots \lambda_{\cal M}, \{\nu\} \rangle = && |\lambda_1, \{\nu_1\} \rangle \otimes | \lambda_2, \{ \nu_2 \} \rangle 
\otimes \ldots  \otimes | \lambda_{\cal M}, \{ \nu_{\cal M} \} \rangle.
\end{eqnarray}
Here, each of the wave functions $| \lambda_{a}, \{ \nu_{a}\}\rangle$ has its support on the $d-$dimensional volume ${\cal R}_a$. As throughout, $\lambda_a$ mark the eigenvalues of the symmetry operators $U_a$. Since $H^{'}$ is local, only terms that are of order ${\cal{O}}(||{\cal{R}}_{a}||) = {\cal{O}}(L^{d})$ in perturbation theory might lift the degeneracy of the eigenstates of Eq. (\ref{tensorp}). 

Such an occurrence indeed transpires in, e.g., the PCM where the ground states of the decoupled horizontal Ising chain Hamiltonian
\begin{eqnarray}
\label{H0chain}
H_0 =
-  \sum_{r} J_{x} \sigma^{x}_{r} \sigma^{x}_{r+
\hat{e}_{x}}
\end{eqnarray}
correspond to spins that are polarized along a uniform  (i.e., the $x$) direction \cite{benoit,DBM}. The local perturbation
\begin{eqnarray}
H^{'}=
-  \sum_{r} J_{y} \sigma^{y}_{r} \sigma^{y}_{r+
\hat{e}_{y}}
\end{eqnarray}
flips pairs of spins on neighboring horizontal chains. Only if ${\cal{O}}(L)$ such spins are flipped may one ground state be linked to another. Indeed, for the PCM, the perturbative corrections due to $V$ are exponentially small in $L$ \cite{benoit,DBM}. We consider the Hamiltonian of Eq. (\ref{H0V}) to be defined on a closed cylinder (so that it becomes the cylindrical Hamiltonian $H_{\sf cyl}$). In this case, perturbation theory would suggest that, in the common eigenstates of this Hamiltonian and all of the symmetry operators, the expectation value of $H^{'}$ (and any of the local terms forming $H^{'}$) is exponentially small in $L^{d}$. In particular, this would imply that $H_{\sf zip}$ may only lift the degeneracy of $H_{\sf open}$ by exponentially small contributions that vanish in the thermodynamic limit. Thus, in the thermodynamic limit, the degeneracy may remain to all orders in perturbation theory. 

We note that in the $H^{'} \to 0$ limit of the planar compass model, common eigenstates of the Hamiltonian $H=H_{0}$ and of all of the symmetries  $\{\hat{O}^{y}_{C_{\ell}}\}$ are rather trivial to write down. These are formed by having, along each row, one of the two equal amplitude (i.e., symmetric or antisymmetric) superpositions of all of the spins fully polarized to the right or to the left, i.e., by direct products of states of the form
$\frac{1}{\sqrt{2}} \Big(| \rightarrow \rightarrow \ldots \rightarrow \rangle \pm |\leftarrow \leftarrow \ldots \leftarrow \rangle \Big)$ along each one of the horizontal rows. 
By being ``fully polarized to the right or to the left'', we allude here to all the spins 
being in the $(+1)$ eigenstate ($|\rightarrow \rightarrow \ldots \rightarrow \rangle$) or all of the spins being in the $(-1)$ eigenstate ($ |\leftarrow \leftarrow \ldots \leftarrow \rangle$) of all local Pauli operators $\sigma_r^x$ that are associated with sites $r$ belonging to that row. 

\section{Variational states of the Planar Compass Model}
\label{varsec}

We next label a finite system ground state of $H_{\sf cyl}$ in the common eigenbasis of the symmetries ${\hat{O}}^{y}_{\ell'}$ by $| \psi_{\sf cyl} \rangle$. With, for each $\ell'$, the below $n_{\ell'} = \pm 1$, there are $2^{L}$ states 
\begin{eqnarray}
\label{stringstates}
({\hat{O}}^{x}_{1})^{n_1} ({\hat{O}}^{x}_{2})^{n_2} \ldots ({\hat{O}}^{x}_{L})^{n_{L}} | \psi_{\sf cyl} \rangle,
\end{eqnarray}
that are formed by acting unitarily on the ground state
$| \psi_{\sf cyl} \rangle$. These states correspond to different eigenvalue strings $\lambda_1 \ldots \lambda_L$ of the symmetry operators ${\hat{O}}^{y}_{\ell'}$ and are thus orthogonal to one another. Similar to boundary terms that are imposed when spontaneous symmetry breaking does not occur (as in, e.g., a stack of $L$ decoupled one-dimensional Ising chains at finite temperature), the $2^L$ different symmetry related states of Eq. (\ref{stringstates}) and associated low energy excitations about these may not all share the same energy yet still exhibit identical correlations and appear to be equal probable in the bulk. 
Beyond the viable absence of spontaneous symmetry breaking as ascertained by the bulk states, the discussion in Appendix \ref{sec-high-heuristic} suggests that within the ground state sector of $H_{\sf cyl}$ of the planar compass model, the expectation value of $\langle H^{'} \rangle$ is uniform (being exponentially small in $L$).  If the energy of the vertical bonds  (captured by $H^{'}$) in the state $| \psi_{\sf cyl} \rangle$ vanishes in the thermodynamic limit then, in that limit, the $2^L$ states of Eq. (\ref{stringstates}) become degenerate with the ground state. Indeed, as noted above, consistently with the considerations of Section \ref{sec-high-heuristic}
and Ref. \cite{benoit,DBM}, the expectation values of vertical bonds in $| \psi_{\sf cyl} \rangle$ may indeed be exponentially small in the system size. Putting all of the pieces together, the $2^L$ orthogonal states of Eq. (\ref{stringstates}) may indeed be asymptotically degenerate with the ground state in the thermodynamic limit. Of course, if perturbation theory applies for ratios $J_{y}/J_{x}$ of order unity (as suggested by \cite{benoit,DBM}) then the states of Eq. (\ref{stringstates}) are the $2^L$ direct products of the ground states of each of the $L$ horizontal Ising chains appearing in Eq. (\ref{H0chain}). In the $(J_y/J_x) \to 0$ limit discussed in Section \ref{sec-high-heuristic}, the states of Eq. (\ref{stringstates}) are eigenstates of both $H_{0}$ and all of the symmetry operators $\{\hat{O}^{y}_{C_{\ell}}\}$. Within such states, the energy of the vertical bonds ($H^{'}$) trivially vanishes \cite{explain1dstring}. If, for systems of finite size $L$, the energy $\langle H^{'} \rangle$ is negative then of all of the states of  Eq. (\ref{stringstates}), the state of maximal energy would be the one in which all vertical bonds on one of the two diagonal boundaries of the ($\theta = \pi/4$) parallelogram of Fig. \ref{fig:cylinder} are flipped- corresponding to eigenvalues $\lambda_{\ell}$ that are completely staggered relative to those in the ground state $|\psi_{\sf cyl} \rangle$. These trends associated with our variational states are  similar to those found earlier. Indeed, numerically \cite{DBM}, it was seen the finite size system ground states corresponded to a uniform value of $\lambda_{\ell}$ while the highest energy states corresponded to a maximally staggered values (i.e., $\lambda_{\ell} = \pm (-1)^{\ell}$ along consecutive rows $C^{y}_{\ell}$).


\begin{thebibliography}{28}

\bibitem{LSM} 
E. Lieb, T. Schultz, and D. Mattis,  Ann. Phys. (N.Y.) {\bf 126}, 407 (1961).

\bibitem{Oshikawa} 
M. Oshikawa, Phys. Rev. Lett. {\bf 84}, 1535 (2000).

\bibitem{Matt} 
M. B. Hastings, Phys. Rev. B 69, 104431 (2004).

\bibitem{DMH}
O. Dubinkin, J. May-Mann, and T. L. Hughes, 
Phys. Rev. B {\bf 103}, 125133 (2021).

\bibitem{NBCB} 
Z. Nussinov, M. Biskup, L. Chayes, and J. van den Brink,
Europhys Lett. {\bf 67}, 990 (2004). 

\bibitem{BCN} 
M. Biskup, L. Chayes, and Z. Nussinov,
Commun. Math. Phys. {\bf 255}, 253 (2005).

\bibitem{Ma}
A. Mishra, M. Ma, F.-C. Zhang, S. Guertler, L.-H. Tang, and S. Wan,
 Phys. Rev. Lett. {\bf 93}, 207201 (2004).

\bibitem{NF}
Z. Nussinov and E. Fradkin,  Phys. Rev. B {\bf 71}, 195120 (2005).


\bibitem{benoit}
B. Doucot, M. V. Feigelman, L. B. Ioffe, and A. S. Ioselevich, Phys. Rev. B {\bf 71}, 024505 (2005).

\bibitem{DBM}
J. Dorier, F. Becca, and F. Mila
Phys. Rev. B {\bf 72}, 024448 (2005).

\bibitem{BN}
C. D. Batista and Z. Nussinov,  Phys. Rev. B {\bf 72}, 045137 (2005).

\bibitem{NOC1} 
Z. Nussinov, G. Ortiz, and E. Cobanera, Ann. Phys. (N.Y.) {\bf 327}, 2491 (2012).

\bibitem{NO} 
Z. Nussinov and G. Ortiz,  Phys. Rev. B {\bf 77}, 064302 (2008).

\bibitem{rev-compass}
Z. Nussinov and J. van den Brink,
Rev. Mod. Phys. {\bf 87}, 1 (2015).

\bibitem{MPolarization} 
G. Ortiz and R. M. Martin,  Phys. Rev. B {\bf 49}, 14202 (1994).

\bibitem{AM} 
N. W. Ashcroft, and N. D. Mermin, {\it Solid State Physics} (Saunders College Publishing, Philadelphia, 1976).
	
\bibitem{Long-TQO}
Z. Nussinov and G. Ortiz, Ann. Phys. (N.Y.) {\bf 324}, 977 (2008).

\bibitem{PNAS}
Z. Nussinov and G. Ortiz, Proc. Natl. Acad. Sci. U.S.A. {\bf 106}, 16944 (2009).  

\bibitem{PRLduality}
E. Cobanera, G. Ortiz, and Z. Nussinov, 
Phys. Rev. Lett. {\bf 104}, 020402 (2010).

\bibitem{AIP} 
E. Cobanera, G. Ortiz, and Z. Nussinov,  Adv. Phys. {\bf  60},  679 (2011).

\bibitem{Harris}
A. B. Harris, T. Yildirim, A. Aharony, O. Entin-Wohlman, and I. Ya. Korenblit, 
Phys. Rev. Lett. {\bf 91}, 087206 (2003). 

\bibitem{Arun}
A. Paramekanti, L. Balents, and M. P. A Fisher
Phys. Rev. B {\bf 66}, 054526 (2002). 

\bibitem{LF} 
M. J. Lawler and E. Fradkin, Phys. Rev. B. {\bf 70}, 165310 (2004).

\bibitem{XM} 
C. Xu and J. E. Moore, Phys. Rev. Lett. {\bf 93}, 047003 (2004). 

\bibitem{McGreevyreview}
J. McGreevy,  {\url{https://arxiv.org/pdf/2204.03045.pdf}}
(2022).  

\bibitem{gaiotto} 
D. Gaiotto, A. Kapustin, N. Seiberg, and B. Willett, J. High Energy Phys. {\bf 2015}, 172 (2015).

\bibitem{banks} 
T. Banks and N. Seiberg, Phys. Rev. D {\bf 83}, 084019 (2011). 

\bibitem{QRH} 
M. Qi, L. Radzihovsky, and M. Hermele, Ann. Phys. (N.Y.) {\bf 424}, 168360 (2021). 

\bibitem{son} 
Y.-H. Du, U. Mehta, D. X. Nguyen, and D. T. Son,
{\url{https://arxiv.org/pdf/2103.09826.pdf}} (2021). 

\bibitem{Natis} 
N. Seiberg and S.-H. Shao, SciPost Phys. {\bf 10}, 027 (2021).

\bibitem{johnm}
N. Iqbal and J. McGreevy,
{\url{https://arxiv.org/abs/2106.12610}} (2021). 

\bibitem{Benini}
F. Benini, C. Cordova, and P.-S. Hsin, 
J. High Energy Phys. {\bf 2019}, 118 (2019).

\bibitem{Cordova}
C. Cordova, T. T. Dumitrescu, and K. Intriligator, J. High Energy Phys. {\bf 2019}, 184 (2019).
 
\bibitem{slagle-field-theory}
K. Slagle and Y. B. Kim, Phys. Rev. B {\bf 96}, 195139 (2017).

\bibitem{You} 
Y. You, T. Devakul, F. J. Burnell, and S. L. Sondhi, Phys. Rev. B
{\bf 98}, 035112 (2018).

\bibitem{Stephen}
D. T. Stephen, J. Garre-Rubio, A. Dua, and D. J. Williamson, 
Phys. Rev. Research {\bf 2}, 033331 (2020). 

\bibitem{Shirley3} 
W. Shirley, K. Slagle, and X. Chen, 
SciPost Phys. {\bf 6}, 041 (2019).

\bibitem{tri2020}
T. Devakul, W. Shirley, and J. Wang, Phys. Rev. Research {\bf 2}, 012059(R) (2020). 

\bibitem{hosho}
 Hosho Katsura and Yu Nakayama,
 {\url{https://arxiv.org/pdf/2204.01924.pdf}} (2022). 
 
 \bibitem{cao}
 Weiguang Cao, Masahito Yamazaki, and Yunqin Zheng, 
 {\url{https://arxiv.org/pdf/2206.02727.pdf}} (2022). 
 
 \bibitem{invert1}
 B. Heidenreich, J. McNamara, M. Montero, M. Reece, T. Rudelius, and I. Valenzuela,
 	J. High Energy Phys. {\bf 2021}, 203 (2021).
 
 \bibitem{invert2}
 G. Arias-Tamargo and D. Rodriguez-Gomez,
 {\url{https://arxiv.org/pdf/2204.07523.pdf}} (2022). 
 
 \bibitem{invert3}
 Y. Choi, H. T. Lam, and S.-H. Shao,
 {\url{https://arxiv.org/pdf/2205.05086.pdf}} (2022). 
 
 \bibitem{invert4}
  Lakshya Bhardwaj, Sakura Schafer-Nameki, and Jingxiang Wu, {\url{https://arxiv.org/pdf/2208.05973.pdf}} (2022). 

\bibitem{Elitzur} 
S. Elitzur, Phys. Rev. D. {\bf 12}, 3978 (1975). 

\bibitem{TORIC}
A. Kitaev, Ann. Phys. (N.Y.) {\bf 303}, 2 (2003).

\bibitem{chamon}
C. Chamon,  Phys. Rev. Lett. {\bf 94}, 040402 (2005).

\bibitem{Haah}
J. Haah, Phys. Rev. A {\bf 83}, 042330 (2011).

\bibitem{PhysRevB.94.235157} 
S. Vijay, J. Haah, and Liang Fu, Phys. Rev. B {\bf 94}, 235157 (2016).

\bibitem{zack} 
Z. Weinstein, E. Cobanera, G. Ortiz, and Z. Nussinov, Ann. Phys. (N.Y.) {\bf 412}, 168018 (2020).

\bibitem{zack1} 
Z. Weinstein, G. Ortiz, and Z. Nussinov, Phys. Rev. Lett. {\bf 123}, 230503 (2019). 

\bibitem{lowdeqsmotion} 
 The equations of motion remain invariant also when incorporating external thermal white noise in the equations of motion in the two dual systems (having delta function inner products between the external noise sources at different points in space and time). This is so since (i) the delta function inner products between the external noise sources remain invariant under general unitary transformations
and (ii) dualities are unitary maps \cite{NOC1,PRLduality,AIP}. 
In the thermal equilibria of these dual systems, the dynamics are identical to those in conventional thermal low dimensional short range systems. Since the finite temperature autocorrelation times in these low dimensional systems, are finite and do not scale with the system size, it follows that due to the  dimensional reduction (that is not only that associated with bounds but can also be made exact via dualities), many of the above stabilizer systems might not be immune to thermal fluctuations \cite{NO,zack,zack1}. Thus, some of these exactly solvable systems may not display glassy dynamics with long time correlations. Other theories, however, may exhibit glassy dynamics. Indeed, qualitatively, higher symmetries may naturally lead to a multitude of numerous degenerate and metastable states leading to non-uniform spatial structures and glassy dynamics. Proving the existence of such numerous degenerate states is indeed the central result of the current work. 

\bibitem{lowdt2} While the dynamics of the Chamon, Haah, and similar models that are dual to simple classical Ising systems may be trivial \cite{NO,zack,zack1}, we note that if one places additional restrictions on the allowed dynamics then glassy dynamics will arise. For instance, if in the classical plaquette models \cite{plaq1,plaq2,plaq3}, only single spin flips are allowed then glassy dynamics will appear. The pattern of low energy single defect motion in the plaquette models is very similar to those in the later models of Chamon and Haah's code. In the triangular plaquette models, each defect leads to ``Pascal's triangle" of flipped spins in its wake and these triangles can be an integer power of two. That the dynamics is of the classical plaquette models is glassy is a consequence of constraining the system so as to only allow single spin flips.  With open boundaries, the classical plaquette models \cite{plaq1,plaq2,plaq3} also have a number of independent spins (once a duality mapping is performed to the plaquette variables) that is linear in the system perimeter. This linear scaling is similar to the number of independent symmetries in the Planar Compass Model (PCM) of Section \ref{sec:POC}. Such single spin flip dynamics are allowed by the transverse field ($h$) term in the Xu-Moore plaquette model \cite{XM} of Eq. (\ref{eq:XM}) in its $\sigma^{z}$ basis. Since the Xu-Moore model is dual \cite{NF,PRLduality,AIP} to the PCM, it exhibits thermodynamic and dynamic behaviors identical to those of the PCM (including its phase transition) \cite{rev-compass}. See Ref. \cite{zhoufrank} for insightful detailed analysis of quantum variants of classical plaquette models from a different perspective.



\bibitem{seiberg2021}
N. Seiberg and S.-H. Shao, SciPost Phys. {\bf 10}, 003 (2021).

\bibitem{Pra}
P. Gorantla, H. T. Lam, N. Seiberg, and S.-H. Shao,
Phys. Rev. B {\bf 103}, 205116 (2021).

\bibitem{Bravyi}
S. Bravyi, B. Leemhuis, and B. M. Terhal, Ann. Phys. (N.Y.) {\bf 326}, 839 (2011).


\bibitem{Shirley1} 
W. Shirley, K. Slagle, and X. Chen, Phys. Rev. B {\bf 99}, 115123 (2019). 

\bibitem{PhysRevB.95.155133} 
A. Prem, J. Haah, and R.  Nandkishore, 
Phys. Rev. B {\bf 95}, 155133 (2017).

\bibitem{Nan1} 
R. Nandkishore and M. Hermele,
Annu. Rev. Condens. Matter Phys. {\bf 10}, 295 (2019).
  
\bibitem{Halasz} 
G. B. Halasz, T. H. Hsieh, and L. Balents,  Phys. Rev. Lett. {\bf 119}, 257202 (2017). 
   
\bibitem{Shirley2}  
W. Shirley, K. Slagle, and X. Chen,
   Ann. Phys. (N.Y.) {\bf 410}, 167922 (2019). 

\bibitem{Shirley4} 
W. Shirley, K. Slagle, Z. Wang, and X. Chen, Phys. Rev. X {\bf 8}, 031051 (2018). 

\bibitem{Shirley5} 
W. Shirley, K. Slagle, and X. Chen,
SciPost Phys. {\bf 6}, 015 (2019).

\bibitem{Slagle1} 
K. Slagle and Yong Baek Kim,  Phys. Rev. B {\bf 97}, 165106 (2018). 

\bibitem{elastic-fracton} 
M. Pretko and L. Radzihovsky,
Phys. Rev. Lett. {\bf 120}, 195301 (2018). 

 \bibitem{Sid2} 
 H. Ma, S. A. Parameswaran, M. Hermele, and R. M. Nandkishore, 
Phys. Rev. B {\bf 97}, 125101 (2018). 

\bibitem{devakul} 
T. Devahul, S. A. Parameswaran, and S. L. Sondhi,  Phys. Rev. B {\bf 97}, 041110 (2018). 

\bibitem{Slagle17}
K. Slagle and Y.-B. Kim, Phys. Rev. B {\bf 96}, 195139 (2017). 

\bibitem{hybrid1} 
N. Tantivasadakarn, W. Ji, and S. Vijay,
Phys. Rev. B {\bf 103}, 245136 (2021). 

\bibitem{hybrid2} 
P.-S. Hsin and K. Slagle, {\url{https://arxiv.org/pdf/2105.09363.pdf}} (2021). 

\bibitem{sala} 
P. Sala, T. Rakovszky, R. Verresen, M. Knap, and F. Pollmann, Phys. Rev. X {\bf 10}, 011047 (2020).

\bibitem{KK}
K. I. Kugel and D. I. Khomskii, Sov. Phys. JETP {\bf 37}, 725 (1973); Sov. Phys. Usp. {\bf 25}, 231
(1982).

\bibitem{jan1}
L. F. Feiner, A. M. Oles, and J. Zaanen, Phys. Rev. Lett. {\bf 78}, 2799 (1997). 

\bibitem{120compass}
J. van den Brink, P. Horsch, F. Mack, and A. M. Oles,
Phys. Rev. B {\bf 59}, 6795 (1999). 

\bibitem{simon}
A. van Rynbach, S. Todo, and S. Trebst,
Phys. Rev. Lett. {\bf 105}, 146402 (2010).

\bibitem{NAG}
Z. Nussinov, Phys. Rev. B {\bf 69}, 014208 (2004).

\bibitem{SN}
 S. Sachdev and D. R. Nelson, Phys. Rev. B {\bf 32}, 1480
(1985).

\bibitem{spiral1} 
Z. Nussinov, {\url{https://arxiv.org/pdf/cond-mat/0105253.pdf}} (2001).

\bibitem{spirals}
H. Yan and J. Reuther, {\url{https://arxiv.org/pdf/2112.10676.pdf}} (2021).

\bibitem{Eckart} 
C. Eckart, Rev. Mod. Phys. {\bf 2}, 305 (1930).

\bibitem{elastic1}
V. Cvetkovic, Z. Nussinov , and J. Zaanen,
Philos. Mag. {\bf 86}, 2995 (2006).

\bibitem{elastic2}
Aron J. Beekman, Jaakko Nissinen, Kai Wu, Ke Liu, Robert-Jan Slager, Zohar Nussinov, Vladimir Cvetkovic, and Jan Zaanen, Physics Reports {\bf 683},  1 (2017).

\bibitem{QOBD}
E Rastelli and A Tassi,  J. Phys. C: Solid State Phys. {\bf 20}, L303 (1987). 


\bibitem{QOBD0}
C.  L. Henley,  Phys. Rev. Lett. {\bf 62}, 2056 (1989).

\bibitem{QOBD1}
A. Chubukov, Phys. Rev. Lett. {\bf 69}, 832 (1992). 

\bibitem{QOBD2}
U. Hizi, P. Sharma, and C. L. Henley, Phys. Rev. Lett. {\bf 95},
167203 (2005).

\bibitem{QOBD3}
U. Hizi and C. L. Henley, 
Phys. Rev. B {\bf 73}, 054403 (2006).

\bibitem{QOBD4}
I. Klich, S.-H. Lee, and K. Iida,
Nature Communications
{\bf 5}, 3497
(2014).

\bibitem{Weyl}
H. Weyl,
{\it The Theory of Groups and Quantum Mechanics} (Dover Publications Inc., New York, 1950). 

\bibitem{Wigner1959} 
E. Wigner, {\it Group Theory and Its Application to the Quantum Mechanics of
Atomic Spectra} (Academic Press, New York, 1959).

\bibitem{Morton}
M. Hamermesh, 
{\it Group Theory and its Application to Physical Problems} (Dover Publications Inc., New York, 1989). 

\bibitem{Tinkham}
M. Tinkham, {\it Group Theory and Quantum Mechanics} (Dover Publications Inc., New York, 2003).

\bibitem{sagi}
E. Sagi, Y. Oreg, A. Stern, and B. I. Halperin,
Phys. Rev. B {\bf 91}, 245144 (2015).

\bibitem{explain-A}
We emphasize here that we are discussing the exact degeneracy $g(E)$ associated with a precise given energy $E$
of the full many body system (a macroscopic quantity scaling with the system size) not the energy density (e.g., the energy per particle or energy per unit volume). Throughout the current work, we will spend considerable time discussing boundary effects. While boundary effects trivially do not change the energy density they may, a priori, significantly change the energy $E$ itself.

\bibitem{BatistaOrtiz-2004}
C. D. Batista  and  G. Ortiz, Adv. Phys. {\bf 53}, 1 (2004). 

\bibitem{julienb}
 Julien Barre, David Mukamel, and Stefano Ruffo, Phys. Rev. Lett. {\bf 87}, 030601 (2001). 
 
 \bibitem{AK}
 H. Araki and E. Lieb, Comm. Math. Phys. {\bf 18}, 160 (1970).

\bibitem{Bacon} 
D. Bacon, Phys. Rev. A {\bf 73}, 012340 (2006).

\bibitem{Kitaev06} 
A. Kitaev, Ann. Phys. (N. Y.) {\bf 321}, 2 (2006).

\bibitem{explain-square-compass-relation}
Specifically, if a vertical line $\ell$ and a horizontal line $\ell'$ intersect at a point then we may set, in the convention of Section \ref{sec:exp}, for such a {\it single} vertical line $\ell$, a unique symmetry operator to be $U_{a=1}= {\hat{O}}^{x}_{\ell}$
and its dual to be $V_{a=1} = {\hat{O}}^{y}_{\ell'}$.
These two dual symmetry operators trivially satisfy both conditions of Lemma 1 for this single (${\cal M} =1$) unique pair of dual symmetries. Indeed, for conventional boundary conditions applied to finite size systems \cite{DBM}, the compass model exhibits only a $2^{\cal{M}}=2$ fold degeneracy. For the ``cylindrical cut'' and a large set of other boundary conditions, we may define additional symmetries $U$ and $V$ satisfying conditions (1) and (2) with ${\cal{M}}={\cal{O}}(L)$. 

For standard open rectangular and periodic boundary conditions, the operators ${\hat{O}}^{x}_{\ell_1}$
for any vertical line $\ell_1$ and the product of the symmetries $({\hat{O}}^{y}_{\ell_1'} {\hat{O}}^{x}_{\ell_2'})$
for any two horizontal lines $\ell_1'$ and $\ell_2'$ trivially commute with one another (satisfying the second relation of Eq. (\ref{aa}) for {\it all} $\ell_1, \ell_1',$ and $\ell_2'$) while failing to generate a single non-trivial commutator (the first equality appearing in Eq. (\ref{aa})). This commutation relation follows since at a single common site $i$ the Pauli operators $\sigma^{x_i}$ and $\sigma^{y}_i$ anticommute. Thus, as is the case for conventional boundary conditions, any pair of horizontal lines will intersect a vertical line at two common sites. Hence for these boundary conditions, ${\cal{M}}=1$. Here, one may not set the operators $U$ and $V$ to be single line operators for a larger value of ${\cal{M}}$. This is so since any vertical line and any horizontal line intersect at a single point and thus lead to Eq. (\ref{ooa}) making it impossible to satisfy the second equality in Eq. (\ref{aa}) for ${\cal{M}}>1$. 
\bibitem{Aztec1}
N. Elkies, G. Kuperberg, M. Larsen, and J. Propp,  J. Algebr. Comb. {\bf 1}
 pp. 111?132 and 219?234. (1992). 
 
\bibitem{Aztec2}
W. Jockusch, J. Propp, and P. Shor, arXiv:math/9801068 (1998). 

\bibitem{Aztec3}
D. Romik, Ann. Probab. {\bf 40}, 611 (2012). 

 \bibitem{explain_SPT} Note that the condition invoked in this Lemma (the absence, in the thermodynamic ($L \to \infty$) limit, of correlations between local boundary degrees of freedom in {\it any} eigenstate) is not satisfied for Symmetry Protected Topological phases, e.g., \cite{SPT1,SPT2,SPT3,SPT4}. As a well known example, we note that the ground state manifold of the open AKLT chain \cite{AKLT} is spanned by four orthogonal ground states that each exhibit long range correlations between the boundary spins. Indeed, adding a ``zipper Hamiltonian'' that connects the boundary spins leads to a periodic AKLT chain having a single ground state. 

\bibitem{SPT1}
Z.-C. Gu and Xiao-Gang Wen. Phys. Rev. B
{\bf 80}, 155131 (2009).
 
\bibitem{SPT2}
F. Pollmann, A. M. Turner, E. Berg, and M. Oshikawa, Phys. Rev. B {\bf 81}, 064439 (2010).

\bibitem{SPT3}
X. Chen, Z.-C. Gu, and X.-G. Wen, Phys. Rev. B {\bf 83}, 035107 (2011).

\bibitem{SPT4}
N. Schuch, D. P\'erez-Garcia, and J. I. Cirac, Phys. Rev. B
{\bf 84}, 165139 (2011).

\bibitem{AKLT} 
I. Affleck, T. Kennedy, E. H. Lieb, and H. Tasaki, Commun.
Math. Phys. {\bf 115}, 477 (1988).

\bibitem{ETH1} 
J. M. Deutsch, Phys. Rev. A {\bf 43}, 2046 (1991). 

\bibitem{ETH2}  
M. Srednicki, Phys. Rev. E {\bf 50}, 888 (1994). 

\bibitem{ETH3} 
M. Rigol, V. Dunjko, and M. Olshanii, Nature {\bf 7189}, 854 (2008). 

\bibitem{ETH4} 
F. Borgonovi, F.M. Izrailev, L.F. Santos, and V.G. Zelevinsky, Phys. Rep. {\bf 626}, 1 (2016); 
W. Beugeling, R. Moessner, and M. Haque,
Phys. Rev. E {\bf 91}, 012144 (2015).

\bibitem{compasscorr} 
To provide some physical intuition, we regress to the compass models. For asymptotically large spatial separation $|r-r'|$ between local operators that are coupled in a bilinear form in $H_{\sf zip}$, the expectation values 
\begin{eqnarray}
\langle \phi^{\mu}_{r} \phi^{\nu}_{r'} \rangle \to C^{\mu \nu} + {\cal{O}}(e^{-|r-r'|/\xi}).
\label{sym:cor}
\end{eqnarray}
Here, $C^{\mu \nu}$ is a constant. Eq. (\ref{sym:cor}) applies to states that lie in a sector of fixed energy. Given Eq. (\ref{sym:cor}), as $L \to \infty$, when $L \to \infty$ the spatial distance $|r-r'|$ between any spins that are coupled to each other in Eq. (\ref{zipper}) diverges in the open lattice geometry of Fig. \ref{fig:cylinder}.


\bibitem{MacDonald}
J. K. L. MacDonald, Phys. Rev. {\bf 43}, 830 (1933).

\bibitem{JonesOrtiz}
M. D. Jones, G. Ortiz, and D. M. Ceperley
Phys. Rev. E {\bf 55}, 6202 (1997).

\bibitem{watanabe}
H. Watanabe,
Phys. Rev. B {\bf 98}, 155137 (2018). 

\bibitem{zeromode1}
E. Cobanera, G. Ortiz, and Z. Nussinov, 
Phys. Rev. B {\bf 87}, 041105 (2013).

\bibitem{zeromode2}
P. Fendley, 
J. Phys. A: Math. Theor. {\bf 49}, 30LT01  (2016).
	
\bibitem{Peierls}
R. Peierls, Proc. Cambridge Phil. Soc. {\bf 32}, 477 (1936).
	
\bibitem{Baxter}
R. J. Baxter, {\it Exactly solved Models in Statistical Mechanics}
(Academic Press, London, 1982).

\bibitem{Klein}
Z. Nussinov, C. D. Batista, B. Normand, and S. A. Trugman
Phys. Rev. B {\bf 75}, 094411 (2007).

\bibitem{Klein'}
B. Normand and Z. Nussinov,
Phys. Rev. Lett. 
{\bf 112}, 207202 (2014). 

\bibitem{topoclass}
M-S. Vaezi, G. Ortiz, and Z. Nussinov,
Phys. Rev. B  {\bf 93}, 205112 (2016).


\bibitem{explain1dstring}
The reader should bear in mind that our simple discussion here is only very qualitative in order to convey the point that the difference in the perturbative corrections of $H'$ between different eigenstates are small (e.g., these may be exponentially small in the system size as the numerical results of \cite{DBM} imply). In the extreme $J_y=0$ (i.e., $H^{'} = 0$) limit, any application of a local $\sigma^x_{r}$ on any of the eigenstates of $H_{0}$ will, of course, not alter the energy. The $2^{L}$ common eigenstates of $H_{0}$ and of all of the horizontal symmetries formed by the direct products of 
$\frac{1}{\sqrt{2}} \Big(| \rightarrow \rightarrow \ldots \rightarrow \rangle \pm |\leftarrow \leftarrow \ldots \leftarrow \rangle \Big)$ along each of the horizontal rows will, under the application of any local $ \sigma^x_{r}$ will permute the states amongst themselves. 
At any finite $J_y$, the eigenstates of $H_{\sf cyl}$ display entanglement between spins in different horizontal rows. 

\bibitem{OBD}
J. Villain, R. Bidaux, J. P. Carton, and R. Conte, J. Physique {\bf 41},
1263 (1980).

\bibitem{OBD1}
E. F. Shender, Sov. Phys. JETP {\bf 56}, 178 (1982). 

\bibitem{OBD2}
C. L. Henley, Phys. Rev. Lett. {\bf 62}, 2056 (1989).

\bibitem{OBD3}
R. Moessner, Can.
J. Phys. {\bf 79}, 1283 (2001).

\bibitem{JoergPete}
J. Schmalian and P. G. Wolynes,
Phys. Rev. Lett. {\bf 85}, 836 (2000).

\bibitem{competing}
 Z. Nussinov, I. Vekhter, A. V. Balatsky, Phys. Rev. B {\bf 79}, 165122 (2009).

\bibitem{Fermionization1}
Z. Nussinov, G. Ortiz, and E. Cobanera,
Phys. Rev. B {\bf 86}, 085415 (2012).

\bibitem{oles} 
W. Brzezicki, J. Dziarmaga, and A. M. Oles,
Phys. Rev. B {\sf 75},  134415  (2007).

\bibitem{Wigner1941}
E. P. Wigner, 
Am. J. Math. {\bf 63}, 57 (1941). 

\bibitem{Wigner-Eckart_theorem-General}
 V. K. Agrawala, J. Math. Phys. {\bf 21}, 1562 (1980).

\bibitem{matth}
M. B. Hastings, 
{\url{https://arxiv.org/pdf/2111.01854v1.pdf}} (2021). 

\bibitem{LR}
E. H. Lieb and D. W. Robinson, 
 Commun. Math. Phys. {\bf 28}, 251 (1972).

\bibitem{QCDineq} 
S. Nussinov and M. A. Lampert, Phys. Rep. {\bf 362}, 193 (2002). 


\bibitem{plaq1}
M. E. J. Newman and C. Moore Phys. Rev. E {\bf 60}, 5068 (1999). 

\bibitem{plaq2}
J. P. Garrahan and M. E. J. Newman, Phys. Rev. E {\bf 62} 7670 (2000). 

\bibitem{plaq3}
A. Lipowski, J. Phys. A {\bf 30}, 7365 (1997).

\bibitem{zhoufrank}
Zheng Zhou, Xue-Feng Zhang, Frank Pollmann, and Yizhi You, 
{\url{https://arxiv.org/pdf/2105.05851.pdf}} (2021).



\end{thebibliography}

\end{document}